\documentclass[aps,prd,preprintnumbers,superscriptaddress,nofootinbib,showpacs,twocolumn]{revtex4-1}%
\usepackage{graphicx}
\usepackage{bm,latexsym,amsmath,amssymb,amsfonts,mathrsfs}
\usepackage{ulem}  
\usepackage{dcolumn}   
\usepackage{color}

\newcommand{\simgt}{\lower.5ex\hbox{$\; \buildrel > \over \sim \;$}}
\newcommand{\simlt}{\lower.5ex\hbox{$\; \buildrel < \over \sim \;$}}
\newcommand{\solM}{M_{\odot}}

\usepackage{color}

\begin{document}

\title{Optimal follow-up observations of gravitational wave events with small optical telescopes}


\author{Tatsuya~Narikawa}
\email[Email: ]{narikawa@icrr.u-tokyo.ac.jp}
\affiliation{
Graduate School of Science, Osaka City University, 
Sugimoto-cho 3-3-138, 
Sumiyoshi-ku, Osaka 558-8585, Japan
}
\affiliation{
Institute for Cosmic Ray Research, The University of Tokyo,
Kashiwanoha 5-1-5, Kashiwa, Chiba, 277-8582, Japan
}

\author{Masato~Kaneyama}
\affiliation{
Graduate School of Science, Osaka City University, 
Sugimoto-cho 3-3-138, 
Sumiyoshi-ku, Osaka 558-8585, Japan
}

\author{Hideyuki~Tagoshi}
\affiliation{
Graduate School of Science, Osaka City University, 
Sugimoto-cho 3-3-138, 
Sumiyoshi-ku, Osaka 558-8585, Japan
}
\affiliation{
Institute for Cosmic Ray Research, The University of Tokyo,
Kashiwanoha 5-1-5, Kashiwa, Chiba, 277-8582, Japan
}

\begin{abstract}

We discuss optimal strategy for follow-up observations by 1-3 m class optical/infrared telescopes 
which target optical/infrared counterparts 
of gravitational wave events detected with two laser interferometric gravitational wave detectors. 
The probability maps of transient sources, such as coalescing neutron stars and/or black holes,
determined by two laser interferometers generally spread widely. 
They include the distant region where it is difficult for small aperture telescopes to observe 
the optical/infrared counterparts. 
For small telescopes, there is a possibility that it is more advantageous to search for nearby region 
even if the probability inferred by two gravitational wave detectors is low. 
We show that in the case of the first three events of advanced LIGO, 
the posterior probability map, derived by using a distance prior restricted to a nearby region,
is different from that derived without such restriction. 
This suggests that the optimal strategy for small telescopes to perform follow-up observation 
of LIGO-Virgo's three events are different from what has been searched so far. 
We also show that, when the binary is nearly edge-on, 
it is possible that the true direction is not included in the 90\% posterior probability region. 
We discuss the optimal strategy to perform optical/infrared follow-up observation 
with small aperture telescopes  based on these facts.
\end{abstract}

\date{\today}
\maketitle

\section{Introduction}

After long experimental efforts to construct high sensitivity gravitational wave detectors, 
gravitational wave (GW) signals were finally detected in 2015 during the first observation run of
the advanced LIGO detectors~\cite{Abbott:2016blz,TheLIGOScientific:2016pea}.
Up to now, two GW signals, named GW150914 and GW151226,
have been confirmed to be true signals, 
and one candidate signal, LVT151012, with lower significance was observed. 
All of these signals, including one candidate, are GWs produced by 
the coalescence of binary black holes (BHs). 
The first GW event, GW150914, was produced by the coalescence of BHs with 
masses, $36~M_\odot$ and $29~M_\odot$. 
BHs with such mass have not been observed in x-ray binaries. 
Thus, this observation immediately raised a new astronomical question:  
how these massive stellar mass BHs were formed?
It is now clear that a new era of GW astronomy has started. 

In order to investigate the astrophysical property of GW sources, 
it is very important to determine the direction of the source and to identify the host galaxy. 
However, since the above observations were done with two LIGO detectors, 
the direction of the sources were not determined precisely. 
With two laser interferometers, the source direction is constrained along a ring on the sky
which is determined by the difference of the arrival time of the signal at each detector. 
With additional information of the directional dependence of the antenna pattern of laser interferometers 
and the response of the detectors to two polarizations of GWs, 
the source direction was constrained 
finally to 230 deg$^2$ for GW150914 and 850 deg$^2$ 
for GW151226\footnote{Here, the given sky map area corresponds to the 90\% credible region.}.
However, these accuracies are not enough for optical/infrared telescopes which field-of-view 
are at most only a few square degree. 
In the error region, there are $10^5-10^7$ galaxies 
within the estimated distance of 400 Mpc.
Even in this situation, a lot of electromagnetic (EM) telescopes performed follow-up observation 
of these signals~\cite{Abbott:2016gcq,Abbott:2016iqz,Morokuma:2016hqx,Smartt:2016oeu,Yoshida:2016ddu}. 
Although some of them are 6-8 m class telescopes, 
such like Subaru and DECam, 
most of them are much smaller, 1 m class telescopes. 

Since the first three events were produced by the coalescence of binary BHs, 
it is not surprising that no EM counterparts were observed~\cite{ref:Fermi}.
On the other hand, GWs from neutron star binaries and 
neutron star - black hole binaries are expected to be observed in the near future. 
They are the candidate of the progenitor of the short gamma ray bursts. 
It is also expected that the r-process nucleosynthesis occurs within the matter 
ejected during the coalescence. The ejecta is heated by the decay of the radioactive elements,
and it may be observed in optical/infrared band. This is called kilonova or macronova. 
This radiation is predicted to become about 21 magnitude in optical/infrared band within a few days
when the source is located at about 200 Mpc~\cite{Tanaka:2016sbx}. 

In this paper, we discuss optimal direction for small telescopes to perform follow-up 
observation of GW signals detected with only two laser interferometers. 
First, we reanalyze the three events observed by LIGO O1. 
For small telescopes, it is not easy to observe objects with 21 magnitude. 
Thus, it is possible for small telescopes to detect EM counterparts only if 
the source is located nearby, e.g., within 100 Mpc.  
We evaluate the three dimensional posterior probability distribution for LIGO-Virgo's events
by restricting the distance prior to a nearby region 
where small telescopes can observe the EM counterparts. 
We find that the direction of the most probable region on the sky is very different 
from the cases when no such restriction to the distance prior is applied. 

Next, we discuss the cases when the binary is nearly edge-on. 
We find that if the observation is done only with two LIGO detectors, 
there is a possibility that the 90\% posterior probability map on the sky 
does not include the true location. 
In such a case, the estimated inclination angle and the distance are also very different from the true value. 
This is related to the gravitational wave Malmquist effect
~\cite{Schutz:2011tw,Nissanke:2012dj,Rover:2007ij}\footnote{In \cite{Messenger:2012jy}, 
it is discussed how the selection bias due to imposing a detection threshold is naturally avoided 
by using the full information from the search considering both the selected data and 
ignorance of the data that are thrown away. 
In this work, we do not focus on such selection bias.}. 
Based on these facts, we discuss the strategy to perform follow-up observations
with small telescopes. 

There are already several papers which discussed the strategy to perform the EM follow-up of GW events. 
Among them, in \cite{Salafia:2017ebv}, 
a careful scheduling of the EM follow-up of GW events is proposed 
based on lightcurve models of counterparts
by constructing the detectability map based on the binary's parameters extracted from the GW signal.
One of the differences of this paper from these works is that our work is based on the reanalysis of 
actual GW events observed by LIGO. 
Another difference is we discuss the GW Malmquist bias.

Note that the use of three or more detectors is essential to improve the sky localization accuracy. 
However, even if Virgo \cite{Virgo} as well as KAGRA~\cite{Somiya:2011np, Aso:2013eba} 
join the network of gravitational wave detectors in the near future, 
since each gravitational wave detector has some down time for some commissioning  works, 
there will always be a possibility that only two detectors are operational. 
This paper discusses such cases.

The remainder of this paper is organized as follows.
In Sec.~\ref{sec:O1events}, 
we reanalyze the first three events observed by LIGO O1
and evaluate the posterior probability distribution of the direction 
with the distance prior restricted to the nearby region. 
In Sec.~\ref{sec:Malmquist}, 
we discuss the gravitational wave Malmquist effect. 
In Sec.~\ref{sec:detection_simulation}, 
we discuss the sky localization in the case of nearly edge-on binaries. 
Section~\ref{sec:summary} is devoted to the summary and discussion. 

\section{Effect of distance prior on localization for O1 events}
\label{sec:O1events}

In this section, we evaluate the posterior probability distribution of the direction 
for the first three events of LIGO O1 
by setting the distance prior to the nearby region. 
We use 4096 seconds of data around O1 events available at LIGO Open Science Center~\cite{events}.
We compute probability density functions (PDFs) and model evidences by using stochastic sampling engine 
based on nested sampling~\cite{Skilling:2006,Veitch:2009hd} . 
Specifically, we use the parameter estimation software, \verb|LALINFERENCE| ~\cite{Veitch:2014wba}, 
which is one of the software of LIGO Algorithm Library (LAL) software suite
\footnote{The software is available at http://www.lsc-group.phys.uwm.edu/lal .}.
We follow the analysis done in Ref.~\cite{TheLIGOScientific:2016wfe,TheLIGOScientific:2016pea}. 
However, the calibration error is not taken into account in our analysis.
For simplicity, we use an effective-precessing-spin waveform model 
called \verb|IMRPhenomPv2|~\cite{Hannam:2013oca}, 
in which the waveform is constructed by twisting up the nonprecessing waveform 
with the precessional motion~\cite{Schmidt:2012rh}. 
The \verb|IMRPhenomPv2| has two spin parameters: 
one is an effective inspiral spin $\chi_{\rm eff}=(m_1\chi_1+m_2\chi_2)/(m_1+m_2)$ where $m_1$ and $m_2$ are two component mass, 
$\chi_1$ and $\chi_2$ are the components of the dimensionless spin,
$\chi_i=\mathbf{S}_i\cdot \hat{\mathbf{L}}/m_i^2$, projected along the 
unit Newtonian orbital angular momentum $\hat{\mathbf{L}}$.  $\mathbf{S}_i$ is the spin angular momentum of each star. 
The other spin parameter is an effective precession spin, $\chi_{\rm p}=\rm{max}(A_1S_{1\perp},A_2S_{2\perp})/(A_2m_2^2)$,
where $S_{1\perp}$ and $S_{2\perp}$ are the magnitudes of the projections of two spins in the orbital plane and $A_i=2+(3m_{3-i})/(2m_i)$~\cite{Hannam:2013oca}.

The other parameters which describe the waveform are 
the chirp mass, the mass ratio, 
the luminosity distance to the source,
the right ascension and the declination of the source,
the polarization angle,
the inclination angle which is the angle between the total angular momentum and the line of sight,
the coalescence time, 
the phase at the coalescence time.
We compute the posterior probability distribution for these 11 parameters. 

First, we show results for GW150914 in Fig.~\ref{fig:GW150914}. 
The upper part of Fig.~\ref{fig:GW150914} show the sky map (left) and 
the two dimensional posterior PDF for the luminosity distance 
$d_{\rm L}$ and the inclination $\cos(\theta_{\rm JN})$ (right). 
In this case, the 90\% credible region of the sky map is $\Delta\Omega=193~{\rm deg}^2$. 
Bottom figures in Fig.~\ref{fig:GW150914} show the sky map (left)  
and the two dimensional posterior PDF for the source luminosity distance and the binary inclination, 
obtained with a distance prior restricted to less than 100 Mpc.
The 90\% credible region of the sky map is $616~{\rm deg}^2$. 
The signal-to-noise ratio (SNR) for both cases are 23.9. 

We find that the sky maps for two cases are very different. 
It is shown in the posterior PDF for the source luminosity distance and the binary inclination 
that when no distance prior is applied, 
large distance and $\cos(\theta_{\rm JN})\simeq -1$ (face-off) is favored. 
On the other hand, when the distance prior is applied, 
small distance and $\cos(\theta_{\rm JN})\simeq 0$ (edge-on) is favored.

The same analyses are done for GW151226 and LVT151012. 
Figure~\ref{fig:GW151226} shows sky map and posterior PDF of $d_{\rm L}$-$\cos(\theta_{\rm JN})$ 
for GW151226. 
The 90\% sky localization accuracy is $1032~{\rm deg}^2$ in the case of no strong distance prior 
and $335~{\rm deg}^2$ in the case of the distance prior, $d_{\rm L}\leq 100~{\rm Mpc}$. 
The sky map indicates completely different region along the ring determined by the time delay. 
For the inclination angle, $\cos(\theta_{\rm JN})\simeq \pm 0.8$ is favored 
in the case of no strong distance prior and $\cos(\theta_{\rm JN})\simeq -0.4$ is favored 
in the case of the distance prior, $d_{\rm L}\leq 100~{\rm Mpc}$. 

Figure~\ref{fig:LVT151012} is for LVT151012. 
The 90\% sky localization accuracy is $1415~{\rm deg}^2$ in the case of no strong distance prior 
and $157~{\rm deg}^2$ in the case of a distance prior, $d_{\rm L}\leq 100~{\rm Mpc}$. 
Similar to GW151226, the sky map indicates completely different region along the ring determined by the time delay. 
For the inclination angle, $\cos(\theta_{\rm JN})\simeq \pm 0.8$ is favored 
in the case of no strong distance prior and $\cos(\theta_{\rm JN})\simeq 0.3$ is favored 
in the case of a distance prior, $d_{\rm L}\leq 100~{\rm Mpc}$. 

\section{Relation of origin of the bias to the gravitational Wave Malmquist effect}
\label{sec:Malmquist}
Now, we discuss the reason why the sky maps change.
Among the parameters which describe the waveform, 
five parameters\footnote{We are neglecting the spin effects in these studies.}, including distance $d_{\rm L}$ , inclination angle of the orbital plane $\theta_{\rm JN}$ 
(the angle between the total angular momentum vector and the line of sight), 
polarization angle, and two angle parameters which determine direction of the source on the sky, 
determine the amplitude of GWs together with two mass parameters. 
These five parameters appear in the amplitude of the waveform in the form, 
\begin{eqnarray}
\left[ F_+^2 \left( \frac{1+\cos^2\theta_{\rm JN}}{2} \right)^2 + F_\times^2 \cos^2\theta_{\rm JN} \right]^{1/2} d_{\rm L}^{-1},
\label{eq:amplitude}
\end{eqnarray}
$F_+$ and $F_\times$ are the antenna pattern functions of the detector
which are the functions of the source sky location and the polarization angle. 
This suggests that those parameters are correlated. 

When we evaluate the posterior probability density for a given signal candidate, 
we assume that the source is distributed uniformly in the comoving volume. 
Thus, within a unit distance interval, there are more sources at larger distance. 
This suggests that the posterior probability density of distance becomes larger at larger distance. 
From Eq.~(\ref{eq:amplitude}), we find that for a given amplitude of a signal,
if larger distance is preferred, larger value of $\mid F_+\mid$, $\mid F_\times\mid$ and $\cos(\theta_{\rm JN})$
are also preferred. The sky map is determined in this way. 

When the distance prior is limited to nearby region, in order to realize the same amplitude of 
the signal data,  the sky location with smaller value of 
the antenna pattern functions and $\cos(\theta_{\rm JN})$ are favored.
This produces the sky map which is located along the ring determined by the time delay,
but very different from the original map. 

The origin of this effect is similar to the well-known gravitational wave 
Malmquist effect \cite{Schutz:2011tw,Nissanke:2012dj}, which states that 
gravitational wave detectors detect more face-on/off binaries 
($\cos\theta_{\rm JN} \sim \pm1$) located at relatively larger distance.

\section{Cases for edge-on binary coalescences}
\label{sec:detection_simulation}

\subsection{Cases of O2 sensitivities}

In this section, we discuss the observation of edge-on binary coalescences
with two or three detectors' network of LIGO and Virgo. 
We use the theoretical noise power spectrum density of LIGO and Virgo
in \cite{Aasi:2013wya,Singer:2014qca} which were expected 
to be realized during the second observation (O2) of LIGO. 
The range for BNS is 108~Mpc for LIGO and 36~Mpc for Virgo\footnote{The actual second observation of LIGO started on November 30th 2016.
The range for BNS are around 60 to 80 Mpc.}.
As in the previous section, we perform the parameter estimation with \verb|LALINFERENCE|. 
For simplicity, we use a frequency-domain, spin-less inspiral waveforms, 
\verb|TaylorF2|, ~\cite{Buonanno:2009zt} as both signals and templates. 
The mass of the signals are $(1.4~\solM,~1.4~\solM)$. 

Here, we consider three cases. 
Case A: two detectors' network of LIGO Hanford (H) and LIGO Livingston (L), and no distance prior is used. 
Case B: two detectors' network of LIGO Hanford and LIGO Livingston, and a distance prior is set to $d_{\rm L} \leq 30$ Mpc.
Case C: three detectors' network of LIGO Hanford, LIGO Livingston and Virgo (V), and 
no distance prior is used.

In Fig.~\ref{fig:O2_GoodForVirgo}, we show the results of the parameter estimation
in the case of nearly edge-on ($\theta_{\rm JN}$=75~deg, $\cos(\theta_{\rm JN})\sim 0.26$) and distance of 20~Mpc. 
The right ascension and the declination for the injected signal are 12.1~h and 29.3$^\circ$, 
and the SNR for the injected signal at each detector is 11.3 (H), 12.9 (L) and 13.5 (V).
The upper two figures are for case A, 
the middle two figures are for case B, and 
the bottom two figures are for case C. 
The estimated network SNR, the estimated 90\% sky localization accuracy, and the estimated cosine of the offset angle $\delta$ between 
true location and the location of the maximum PDF are, 
for case A: network SNR=16.3, 735 ${\rm deg}^2$, $\cos(\delta)=-1.00$, 
for case B: network SNR=19.8, 605 ${\rm deg}^2$, $\cos(\delta)=0.960$, 
and for case C: network SNR=22.1, 10.8 ${\rm deg}^2$, $\cos(\delta)=1.00$. 
We find that in the case of the HL network in case A, both the estimated sky location and the estimated inclination angle 
are largely biased.
In this case, larger distance and smaller inclination angle are more favored 
than true value. 
The reason for this is similar to the GW Malmquist effect: 
the region with larger distance have a larger volume which produce 
larger posterior probability density, and consequently smaller inclination angle is favored.
On the other hand, we find in case B, 
if a strong distance prior is applied, the bias of the sky location and the inclination angle 
become much smaller.  However, the sky localization accuracy is still not good. 
Dramatic improvement can be found in case C. Especially, the sky location indicates the correct direction, and 
the sky localization accuracy becomes about 10 ${\rm deg}^2$. 
This clearly show that, even if the sensitivity of third detector is only about 1/3 of other two detectors, 
the presence of the third detector is essential to improve the sky localization accuracy.
Note however that, even in case C, 
the maximum posterior probability density of the inclination angle 
points $\cos(\theta_{\rm JN})\sim 1.00$. 
Corresponding to this, there is a second peak of the posterior probability density of the distance around 38~Mpc, 
although the distance is constrained much better than two detectors' case.

In Fig.~\ref{fig:O2_GoodForHL}, we show an another example of a nearly edge-on binary. 
In this case, the distance and the inclination angle is the same as in Fig.~\ref{fig:O2_GoodForVirgo}, 
but the sky location for the injected signal is different. 
The right ascension and the declination are 4.76~h and 13.4$^\circ$, 
and the SNR at each detector for the injected signal is 23.1 (H), 16.3 (L), and 3.54 (V).
The estimated network SNR, the estimated 90\% sky localization accuracy, and the estimated cosine of the offset angle $\delta$ between 
true location and the location of the maximum PDF are 
for case A: network SNR=29.7, 77.4 ${\rm deg}^2$, $\cos(\delta)=-0.995$, 
for case B: network SNR=27.8, 336  ${\rm deg}^2$, $\cos(\delta)=0.985$, 
and for case C: network SNR=28.1, 31.9 ${\rm deg}^2$, $\cos(\delta)=0.994$. 
We find that in case A, the estimated sky location points completely different direction 
from the true one. 
The estimated inclination angle favors face-off ($\cos(\theta_{\rm JN})\sim -1.00$)
which is very different from the true value. 

We find in case B of Fig.~\ref{fig:O2_GoodForHL} that 
when we apply a strong distance prior ($d_{\rm L}\leq 30$~Mpc), the difference of the sky location and the inclination angle 
become much smaller.  However, the sky localization accuracy is still not good, 
and the posterior distribution of the inclination angle is divided into two region, and 
the accuracy is not very good. 

On the other hand, in case C of Fig.~\ref{fig:O2_GoodForHL}, the sky location indicates the correct direction, 
and the sky localization accuracy is improved dramatically again. 
However even in this case, the accuracy of the inclination angle and the distance are still not good. 
Much larger distance ($\sim 60$~Mpc) than the injected signal and 
the inclination angle of face-on ($\cos(\theta_{\rm JN})\sim 1.00$) are favored.

The above results suggest an important caveat for the strategy of the follow-up observation. 
In the case of two detector network, 
the estimated direction can be completely different from the true one
in the case of edge-on binaries. 
In such cases, it may be difficult to observe EM counterparts 
if we observe only high probability region.

It is effective to evaluate the PDF
by restricting the distance prior to a nearby region in order to improve the sky localization
and the inclination.
However, the accuracy is still not good enough for performing the follow-up observation. 
The best way to improve the situation is to add one more detector. 
The sky location and the sky localization accuracy can be improved dramatically with three detectors. 
However, even in that case, 
there can appear large bias in the estimated distance and the inclination angle. 
If we believe the PDF for distance, and 
search the galaxies located only around 60~Mpc, we may miss the EM counterparts. 
It is thus important to search the large region of the distance including nearby region 
in order not to miss EM counterparts. 

\subsection{Cases of design sensitivities}
Next, we investigate a case of BNS at 200 Mpc 
detected by LIGO and Virgo at their design sensitivity,
which is more realistic in a sense that 
such events are expected to occur frequently in the near future. 
In Fig.~\ref{fig:event10_BNS200Mpci75deg_DesignHLV}, 
we show the results of a nearly edge-on ($\theta_{\rm JN}=75~{\rm deg}$) BNS at 200 Mpc. 
The right ascension and the declination for the injected signal are 3.12~h and -28.9$^\circ$,
and the SNR of the injected signal at each detector are 6.46 (H), 6.92 (L), and 1.07 (V).
We consider a distance prior $d_{\rm L}\leq250~{\rm Mpc}$ for case B.
The estimated network SNR, the estimated 90\% sky localization accuracy, and the estimated cosine 
of the offset angle $\delta$ between 
true location and the location of the maximum PDF are 
for case A: network SNR=10.7,  $1044.3~{\rm deg}^2$, $\cos(\delta)=-0.803$, 
for case B: network SNR=9.82,   $1125.3~{\rm deg}^2$, $\cos(\delta)=0.927$, 
and for case C: network SNR=9.73,  $463.2~{\rm deg}^2$, $\cos(\delta)=0.994$. 
These results are qualitatively the same as in the cases of O2 sensitivities in Fig.~\ref{fig:O2_GoodForHL}. 
In case A of Fig.~\ref{fig:event10_BNS200Mpci75deg_DesignHLV}, 
the estimated sky location spread widely.
The peak of the posterior probability density of distance becomes larger than 
the true location due to ``the GW Malmquist effect.''
In case B of Fig.~\ref{fig:event10_BNS200Mpci75deg_DesignHLV},
when we apply a distance prior ($d_{\rm L}\leq250~{\rm Mpc}$), 
the estimated distance is improved. 
However, the sky localization accuracy is not improved very much. 
In case C of Fig.~\ref{fig:event10_BNS200Mpci75deg_DesignHLV},
the sky localization accuracy is dramatically improved.
However, the peak of the PDF of distance and $\cos(\theta_{JN})$ is maximum around 400~{\rm Mpc}
and 0.7 respectively. Thus, large bias in the estimated distance and inclination angle will occur 
even in the three detector case.

\section{Summary and discussion}
\label{sec:summary}
In this paper, we investigated the optimal direction for 1-3 m class optical/infrared telescopes 
to search for optical/infrared counterparts 
of GW events detected with two laser interferometric gravitational wave detectors. 
The posterior probability distribution of the distance, the inclination angle and the sky location, 
determined with two laser interferometers generally spread widely. 
They include the distant region where it is difficult for small aperture telescopes to detect  
optical/infrared counterparts if we assume a standard picture of the brightness of the optical/infrared counterparts
like kilonova/macronova. 
We showed that in the case of first three events of advanced LIGO, 
the posterior probability maps of the direction on the sky, derived by using a distance prior restricted to a nearby region,
are different from that derived without such restriction. 
It means that the direction which has been observed so far by follow-up observations
may not be optimal direction for small aperture telescopes. 
Optimal direction may be the direction derived with a restricted distance prior. 

Note that  
we discussed only one distance prior of 100 Mpc just to demonstrate a consequence 
of the use of distance prior. 
Of course, the observable range of optical counterparts depends on the aperture of the telescopes. 
Further, to set the abrupt limit to the observable distance for optical telescopes may not be realistic. 
The probability to observe faint objects may gradually decrease as the brightness of the object becomes fainter. 
Thus, to make a more realistic observational strategy, these facts must be taken into account.

Next, we showed two examples of nearly edge-on binary coalescences located nearby. 
In this case, with the two-detector network of LIGO, 
the direction of the maximum PDF is completely different from the true one. 
Since we do not know {\it a priori} the true direction, in this case, it may be difficult to observe to EM counterparts 
if we observe only high probability region. 
The best way to localize the source accurately is to add one more detector, as expected. 
The sky location and the sky localization accuracy can be improved dramatically with three detectors. 
Virgo detector will soon be operational, and KAGRA~\cite{Somiya:2011np, Aso:2013eba} 
will also be operational in a few years. 
So that will be realized in the near future.
However, even in that case, 
large bias in the estimated distance and the inclination angle may remain in some cases. 
If we search for only the region with high posterior probability density for distance, 
we may miss the EM counterparts. 
It is thus important to search the large parameter region of the distance as much as possible  
in order not to miss nearby faint EM counterparts. 

The second observation run of LIGO started on November 30, 2016, 
only with two LIGO detectors. 
It is said that in this observation, three dimensional information of the posterior probability 
distribution in the direction and distance space is provided \cite{Singer:2016eax}. 
With that information, small aperture telescope can observe only nearby region or only nearby galaxies located in 
the high probability region. 
This is one way to optimize the follow-up observation. 

Note also that there is an ambiguity in the model of kilonova/macronova~\cite{Tanaka:2016sbx}. 
Thus, it may be dangerous to decide the follow-up strategy by relying heavily on theoretical models. 
Note also that, as discussed in this paper, in the case of nearly edge-on binaries, 
there is a possibility that the direction of the source is completely different from the true one, 
and it is not possible completely to improve the direction by imposing the distance prior. 
In such a case, we should remember that 
if a region along the ring on the sky determined by the time delay 
has nonzero posterior probability density, it is physically possible that the source is located in that region.
So it is worth performing follow-up observations for a wide region as much as possible.

\section*{Acknowledgment}
T.N. thanks John Veitch, Walter Del Pozzo, Alberto Vecchio 
for very useful explanation of \verb|LALINFERENCE|, and for their hospitality during his stay 
at University of Birmingham.  
We thank Hyung Won Lee, Chunglee Kim, and Jeongcho Kim for a very useful explanation of \verb|LALINFERENCE|, 
and for their hospitality at Inje University. 
We thank Chris van den Broeck for a useful discussion.
This work was supported by MEXT Grant-in-Aid for Scientific Research on Innovative Areas, 
New Developments in Astrophysics Through Multi-Messenger Observations of Gravitational Wave Sources,
Grants No. 24103005, JSPS Core-to-Core Program, A. Advanced Research Networks, 
the joint research program of the Institute for Cosmic Ray Research, University of Tokyo, 
National Research Foundation (NRF) and Computing Infrastructure Project of KISTI-GSDC in Korea. 
This work was also supported by JSPS KAKENHI Grant Numbers, 23540309, 24244028, and 15K05081. 
T. N.'s work was also supported in part by a Grant-in- Aid for JSPS Research Fellows.
This research has made use of data, software and/or web tools obtained from the LIGO Open Science Center 
(https://losc.ligo.org), a service of LIGO Laboratory and the LIGO Scientific Collaboration. 
LIGO is funded by the U.S. National Science Foundation.




\begin{widetext}

\begin{figure}[ht]
\begin{center}
\begin{tabular}{cc}
 \begin{minipage}[b]{0.45\linewidth}
 \begin{center}
   \includegraphics[width=1.0\textwidth,angle=0]{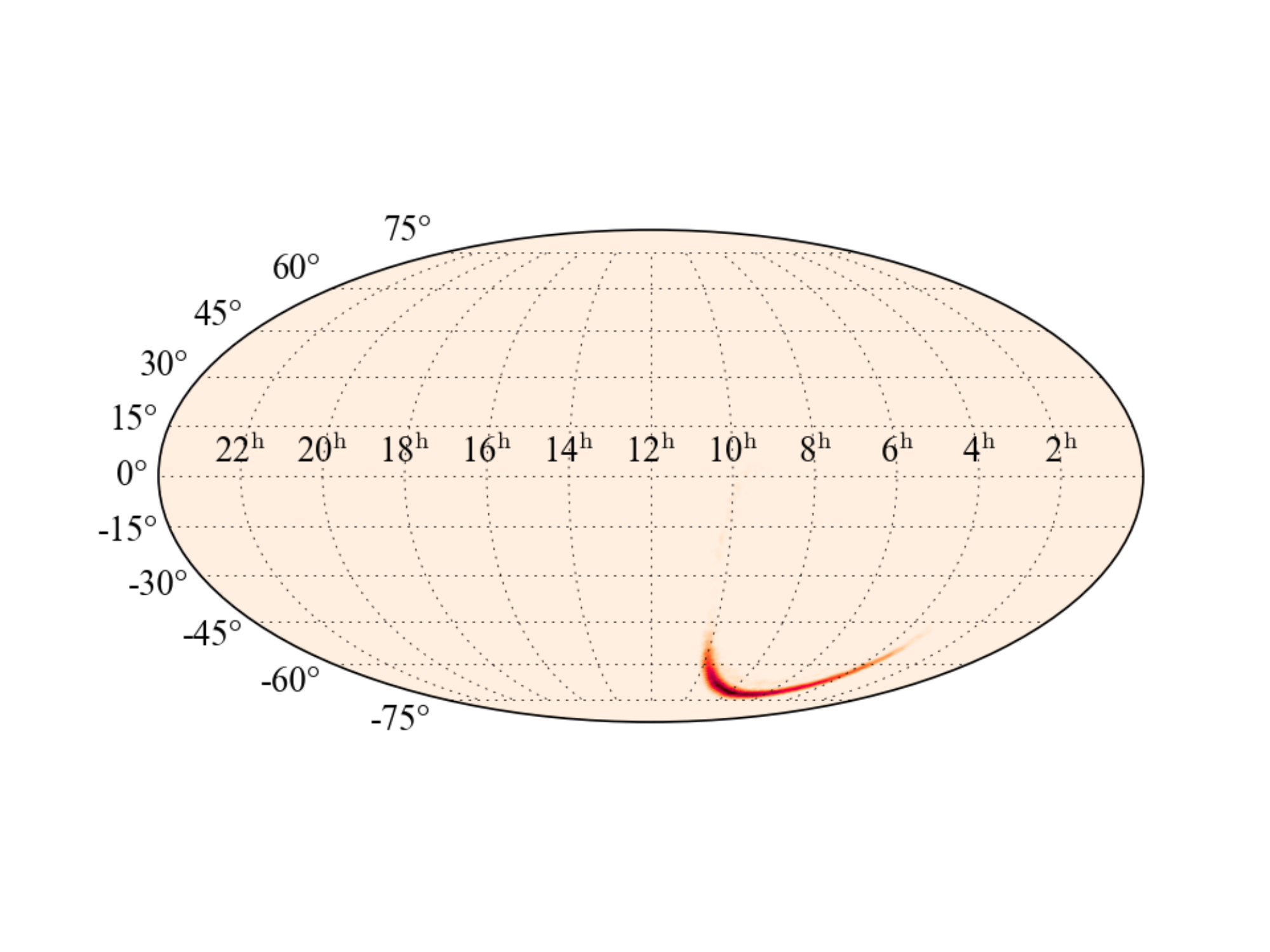}
\end{center}
\end{minipage}
 \begin{minipage}[b]{0.45\linewidth}
 \begin{center}
   \includegraphics[width=1.0\textwidth,angle=0]{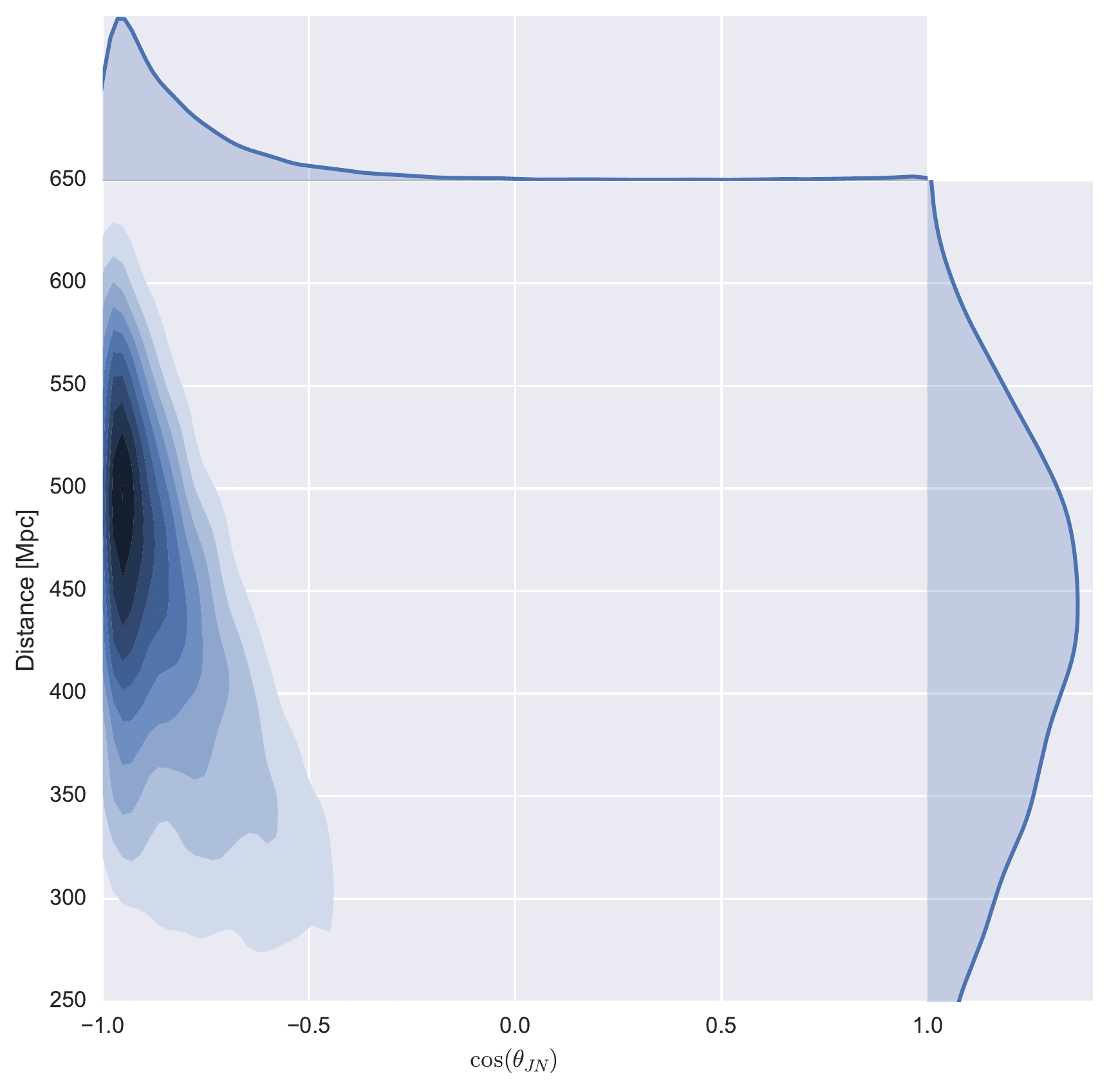}
 \end{center}
 \end{minipage}\\
 \begin{minipage}[b]{0.45\linewidth}
 \begin{center}
   \includegraphics[width=1.0\textwidth,angle=0]{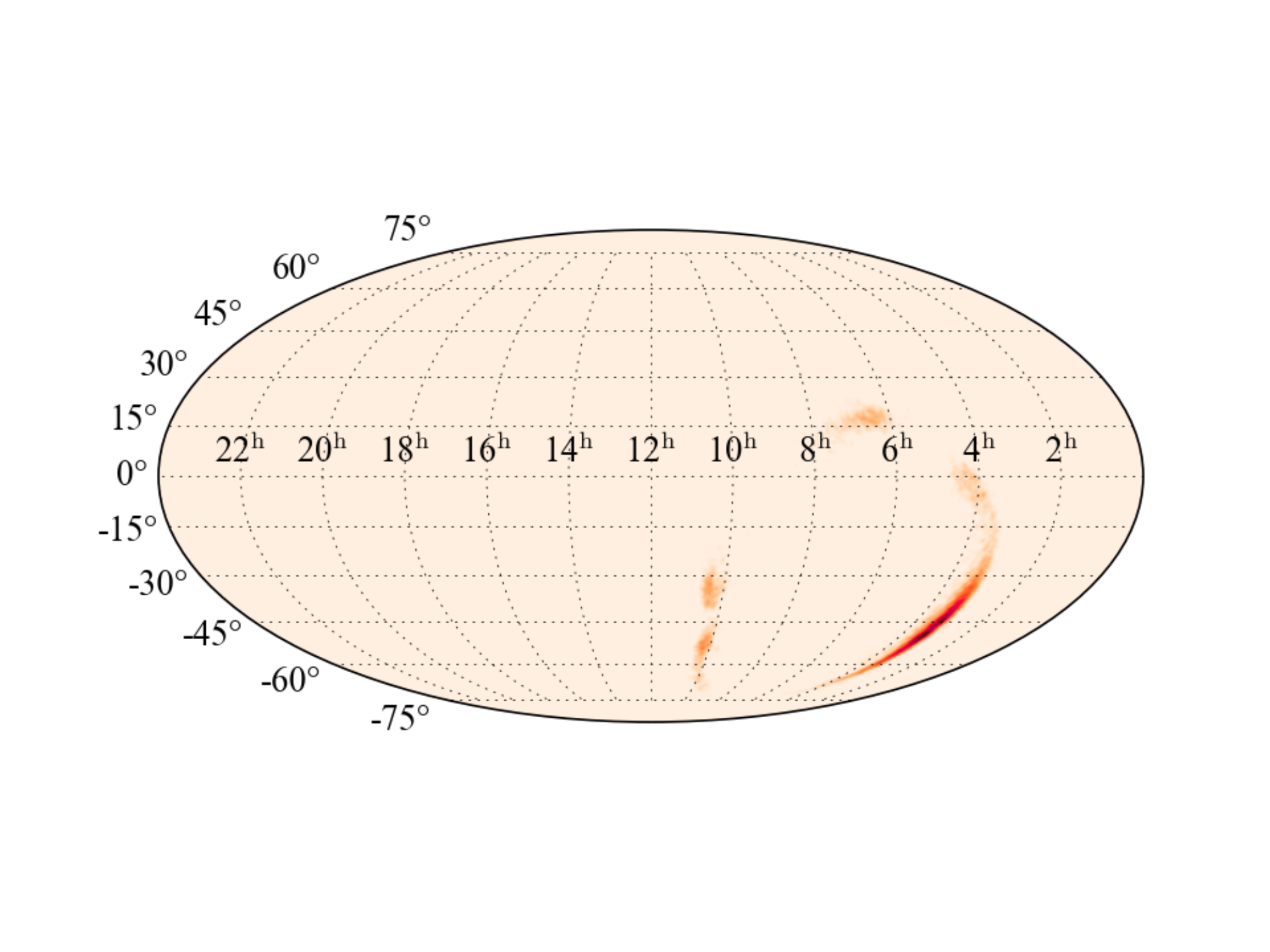}
\end{center}
\end{minipage}
 \begin{minipage}[b]{0.45\linewidth}
 \begin{center}
   \includegraphics[width=1.0\textwidth,angle=0]{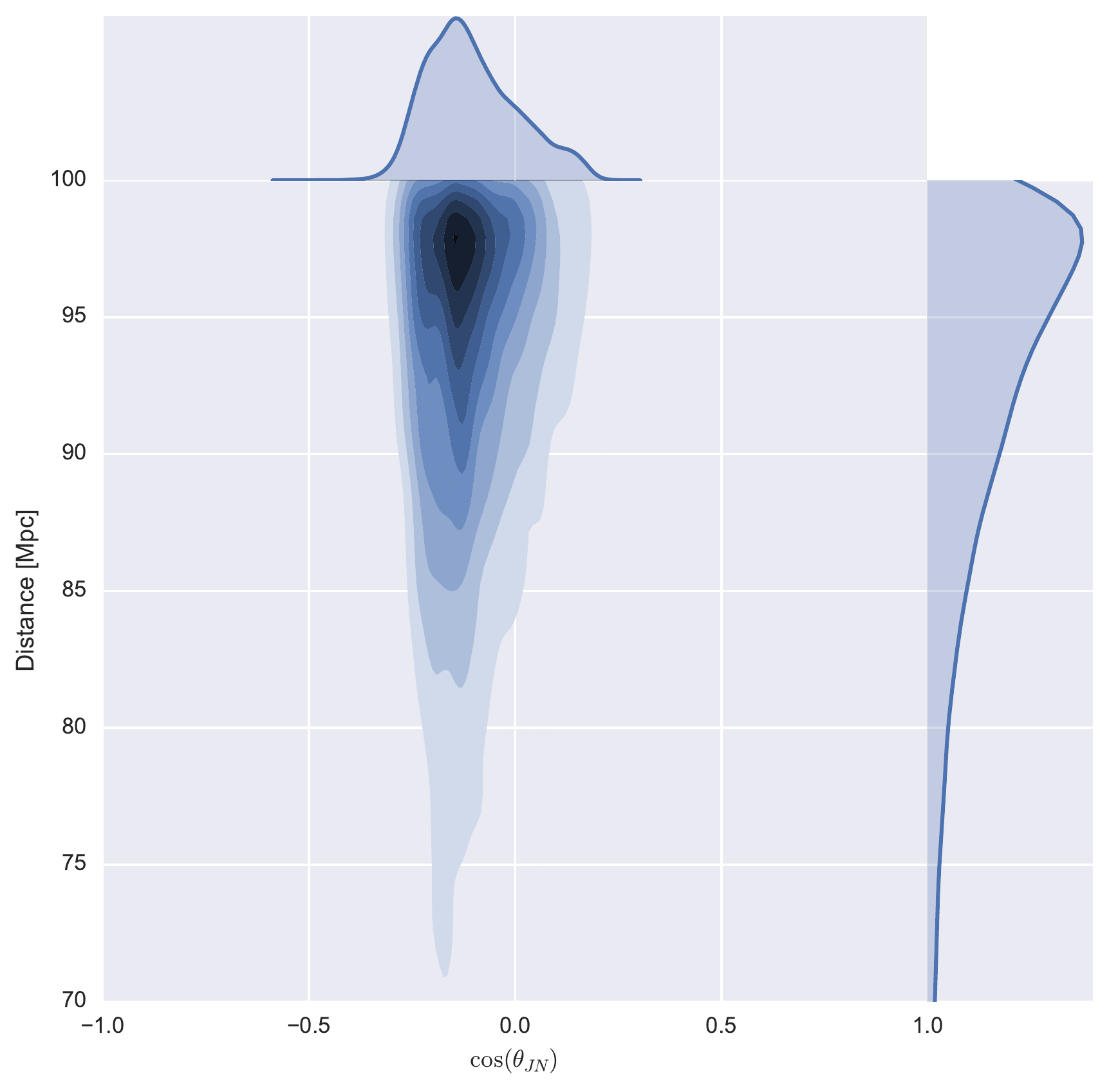}
 \end{center}
 \end{minipage}\\
 \end{tabular}
 \caption{
Posterior PDFs for the sky location and the source luminosity distance and the binary inclination angle for GW150914
when no strong prior is applied (top),
and when a distance prior ($d_{\rm L} \leq 100~{\rm Mpc}$) is used (bottom).
The darker color corresponds to higher probability.}
\label{fig:GW150914}
 \end{center}
\end{figure}
\begin{figure}[ht]
\begin{center}
\begin{tabular}{cc}
 \begin{minipage}[b]{0.45\linewidth}
 \begin{center}
   \includegraphics[width=1.0\textwidth,angle=0]{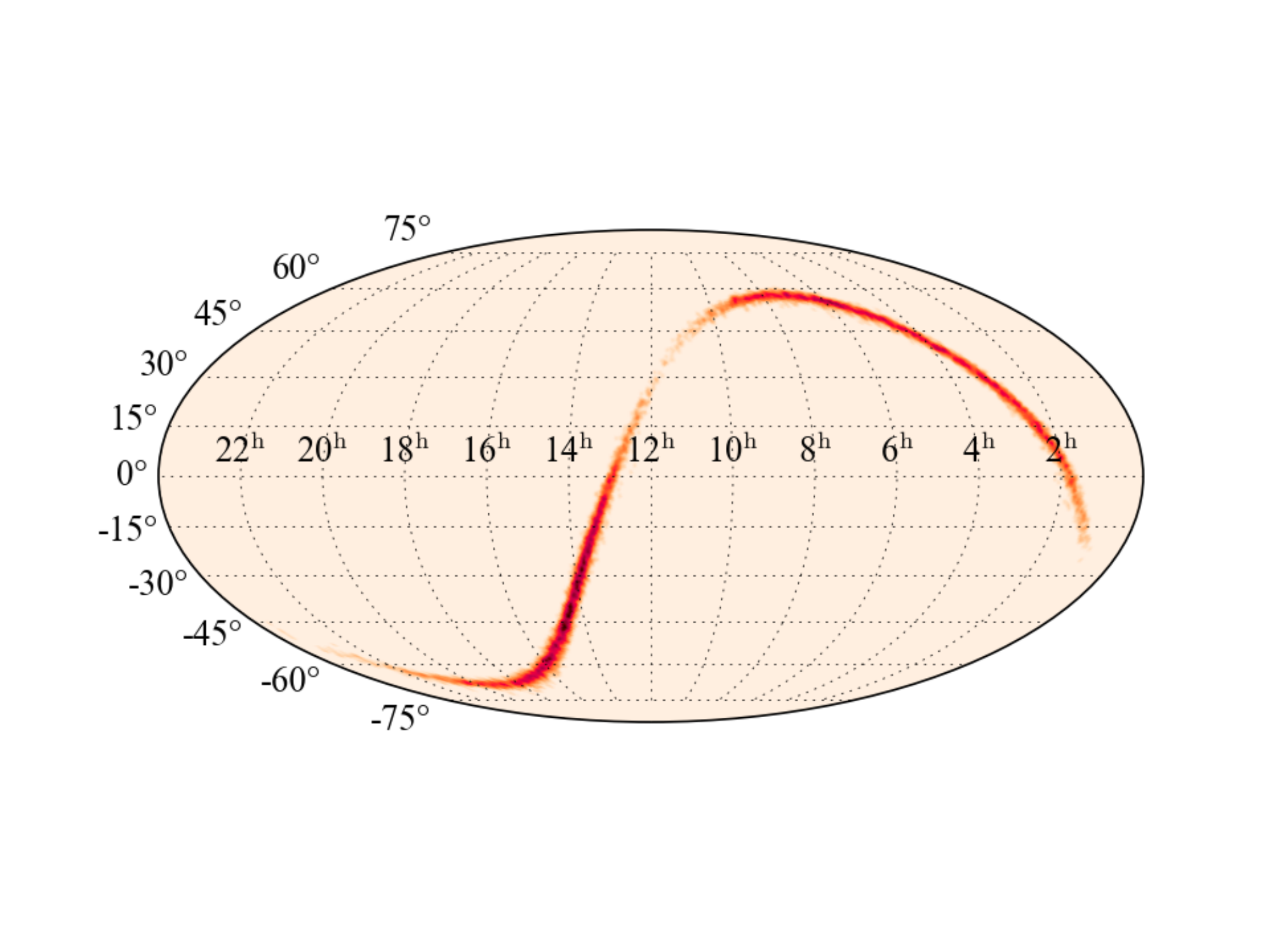}
\end{center}
\end{minipage}
 \begin{minipage}[b]{0.45\linewidth}
 \begin{center}
   \includegraphics[width=1.0\textwidth,angle=0]{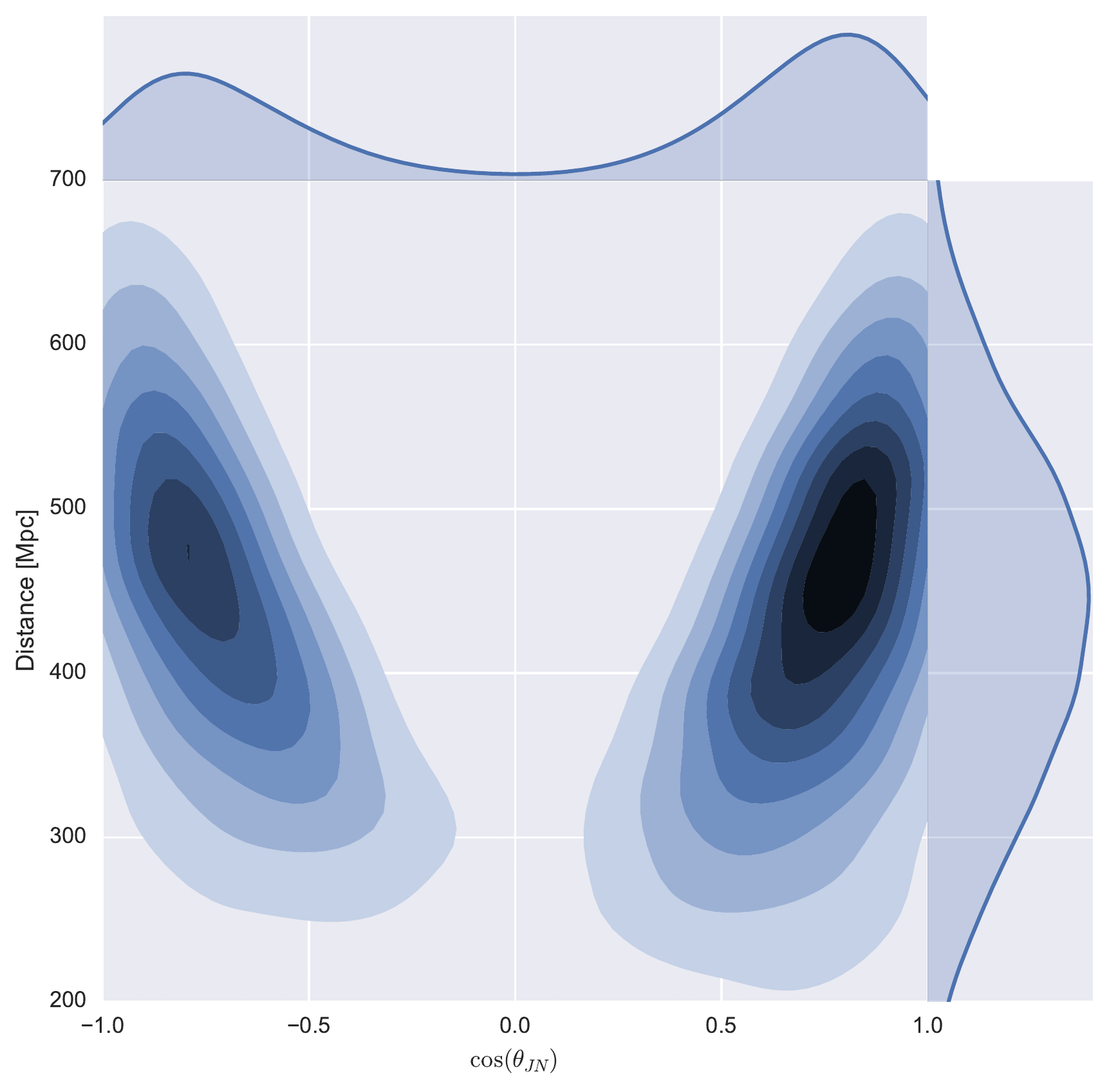}
 \end{center}
 \end{minipage}\\
 \begin{minipage}[b]{0.45\linewidth}
 \begin{center}
   \includegraphics[width=1.0\textwidth,angle=0]{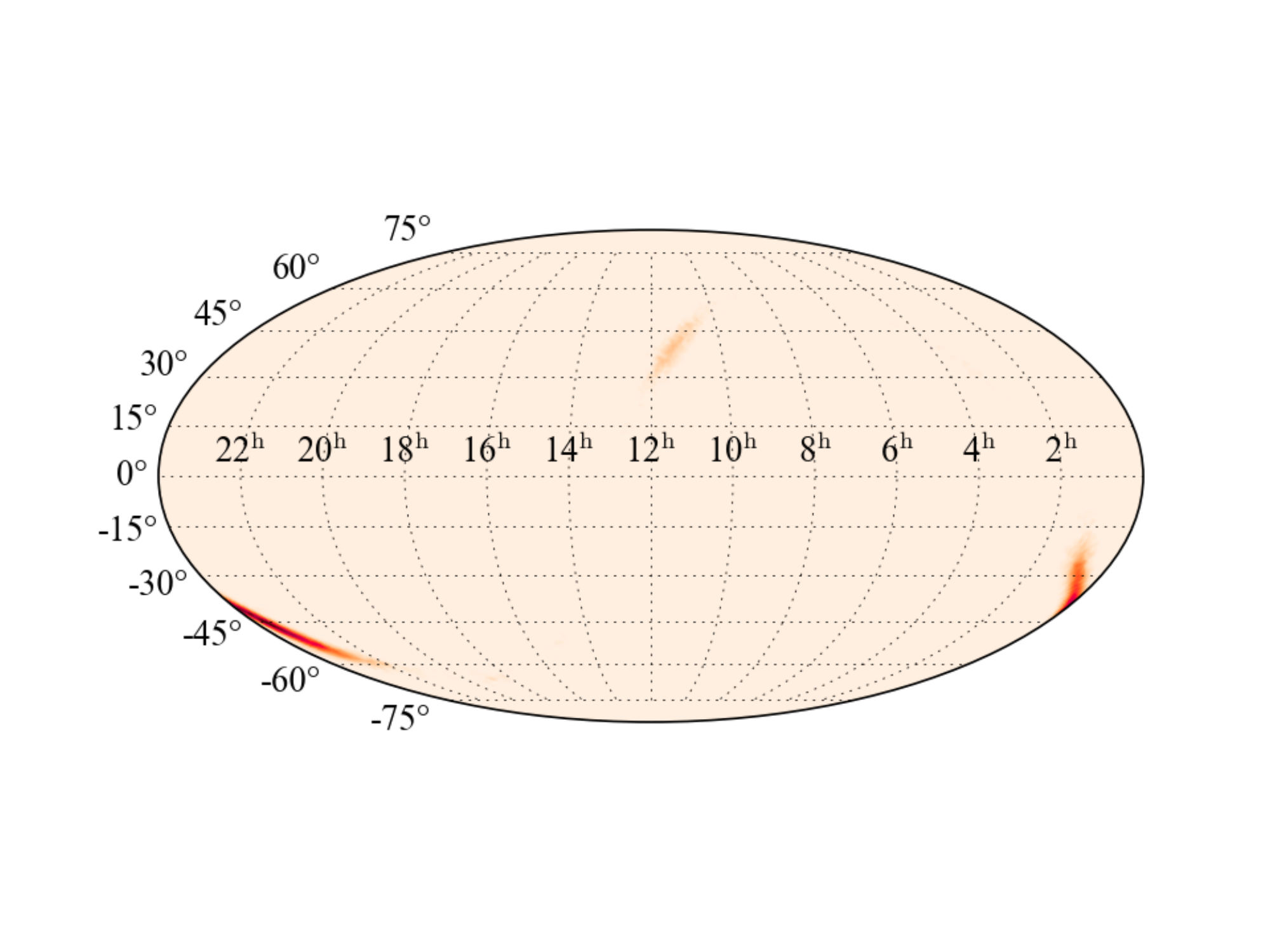}
\end{center}
\end{minipage}
 \begin{minipage}[b]{0.45\linewidth}
 \begin{center}
   \includegraphics[width=1.0\textwidth,angle=0]{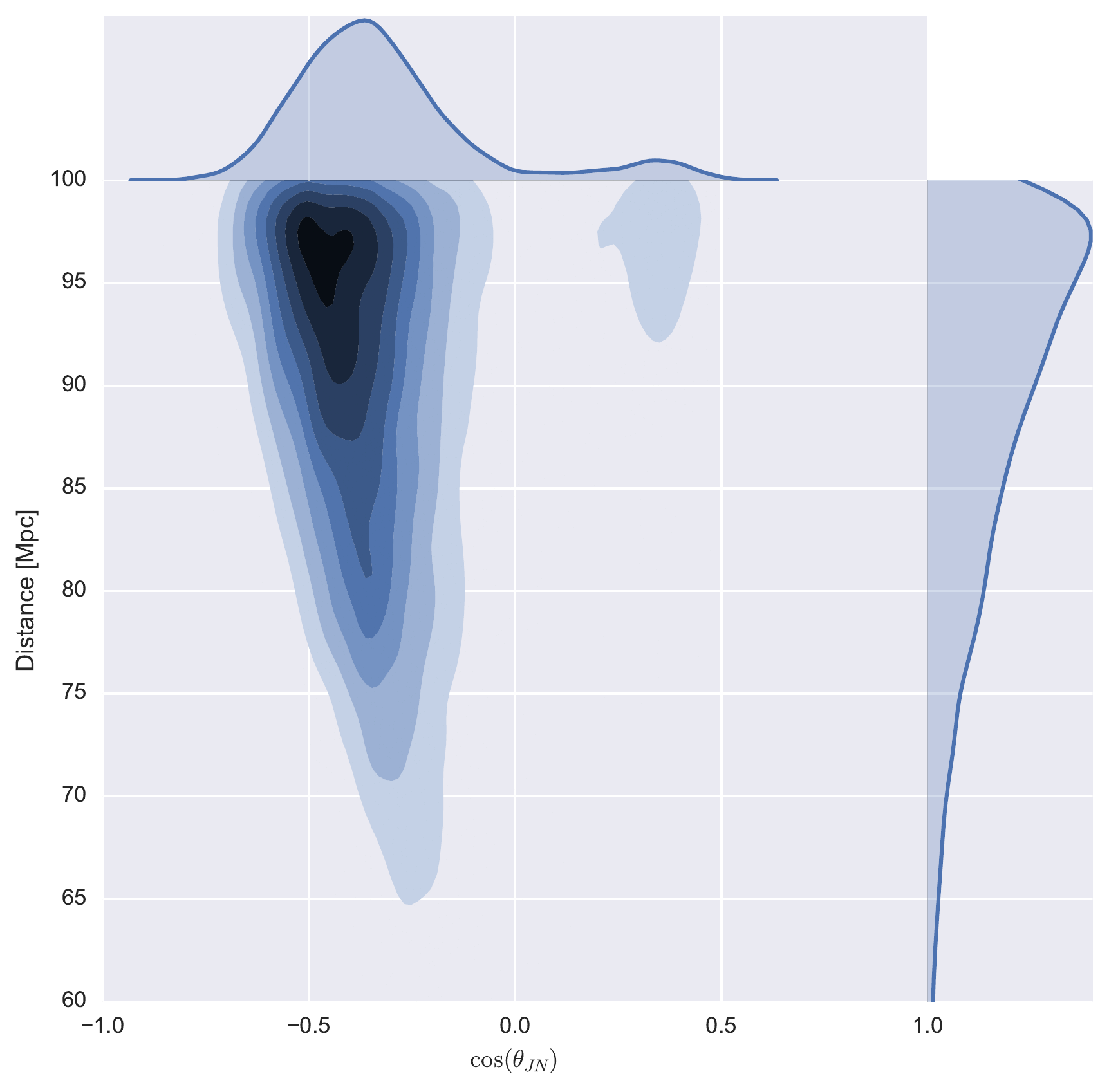}
 \end{center}
 \end{minipage}\\
 \end{tabular}
 \caption{
The same figure as Fig.~\ref{fig:GW150914} but for GW151226.
}
\label{fig:GW151226}
 \end{center}
\end{figure}
\begin{figure}[ht]
\begin{center}
\begin{tabular}{cc}
 \begin{minipage}[b]{0.45\linewidth}
 \begin{center}
   \includegraphics[width=1.0\textwidth,angle=0]{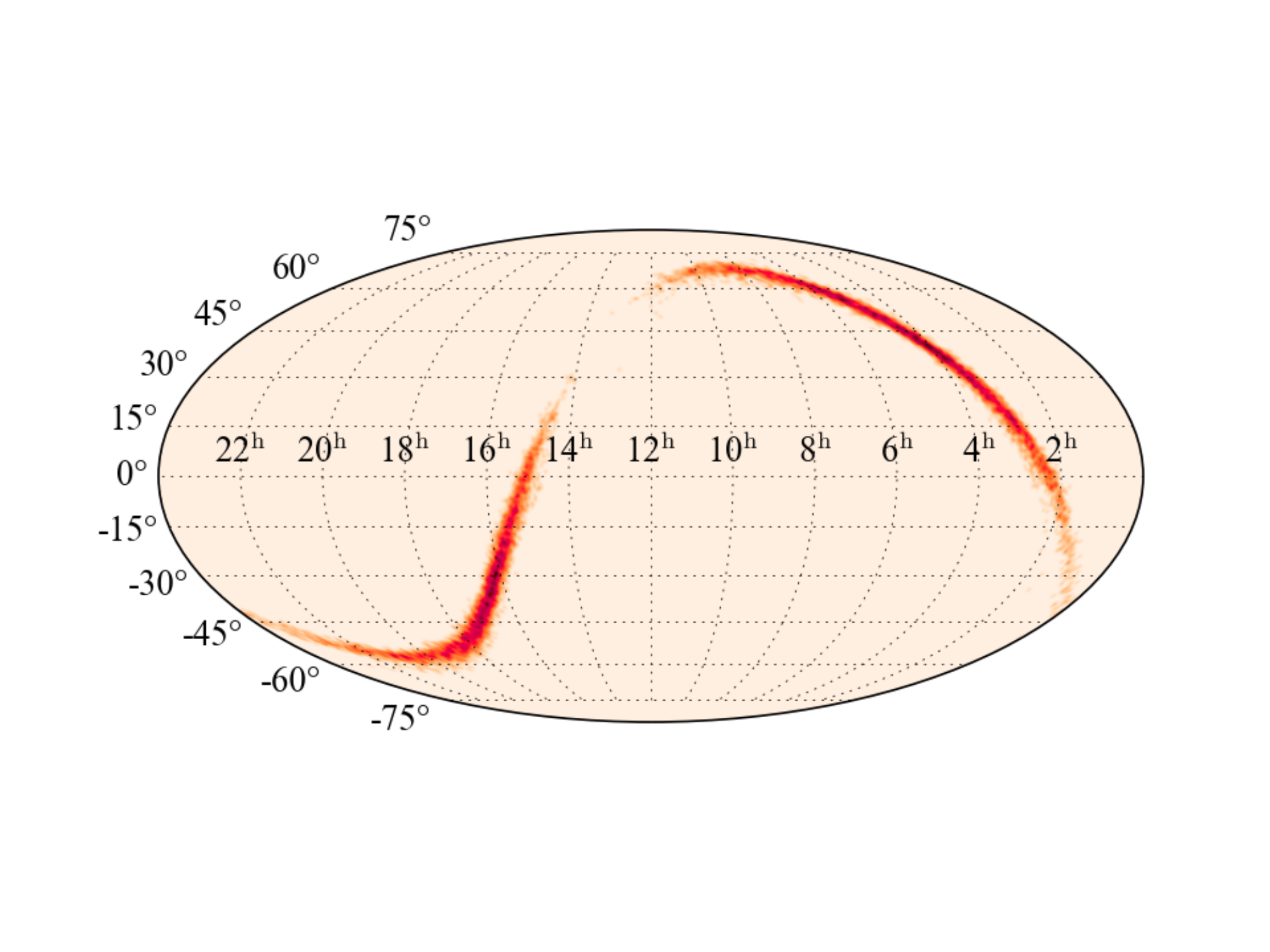}
\end{center}
\end{minipage}
 \begin{minipage}[b]{0.45\linewidth}
 \begin{center}
   \includegraphics[width=1.0\textwidth,angle=0]{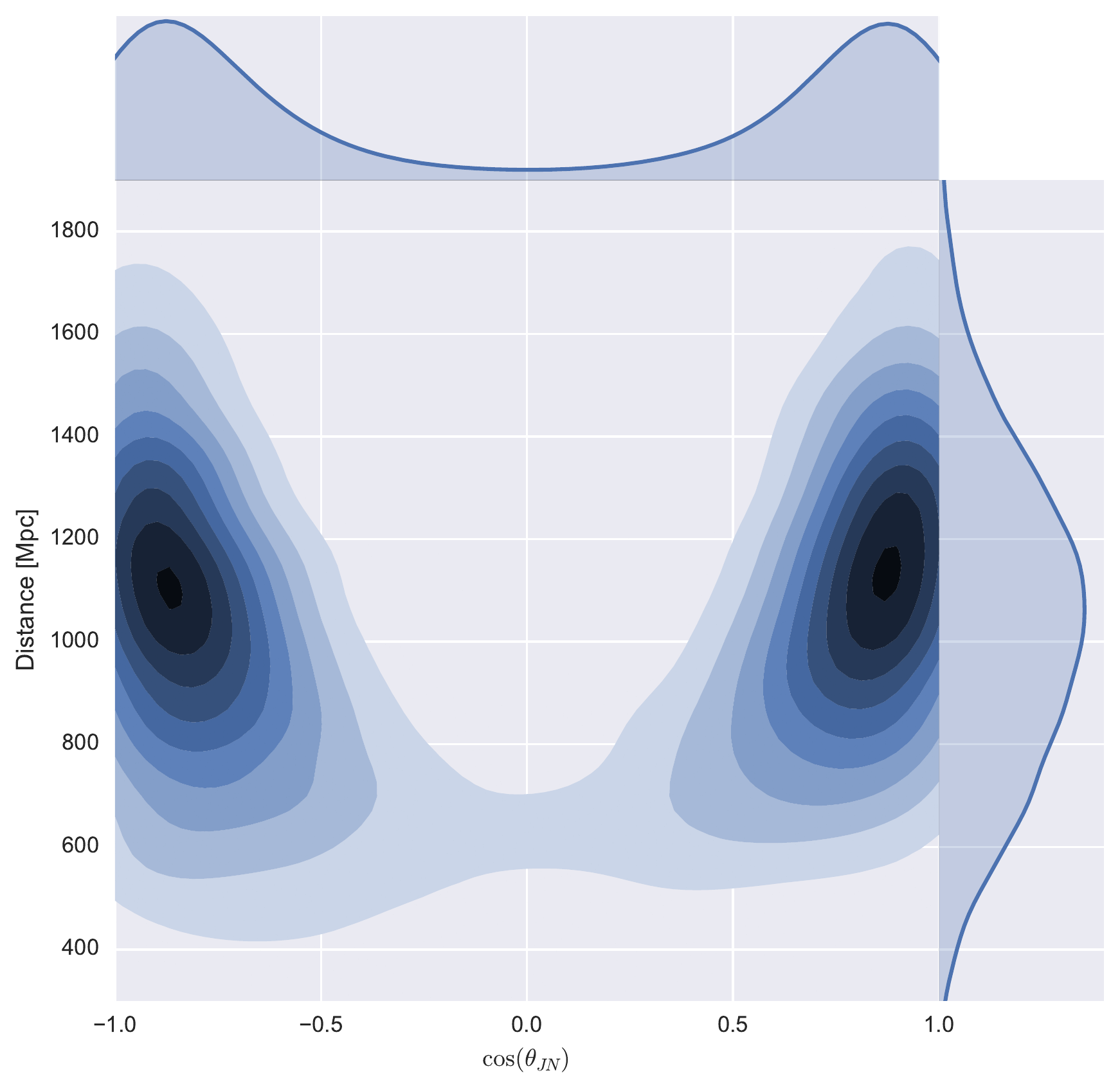}
 \end{center}
 \end{minipage}\\
 \begin{minipage}[b]{0.45\linewidth}
 \begin{center}
   \includegraphics[width=1.0\textwidth,angle=0]{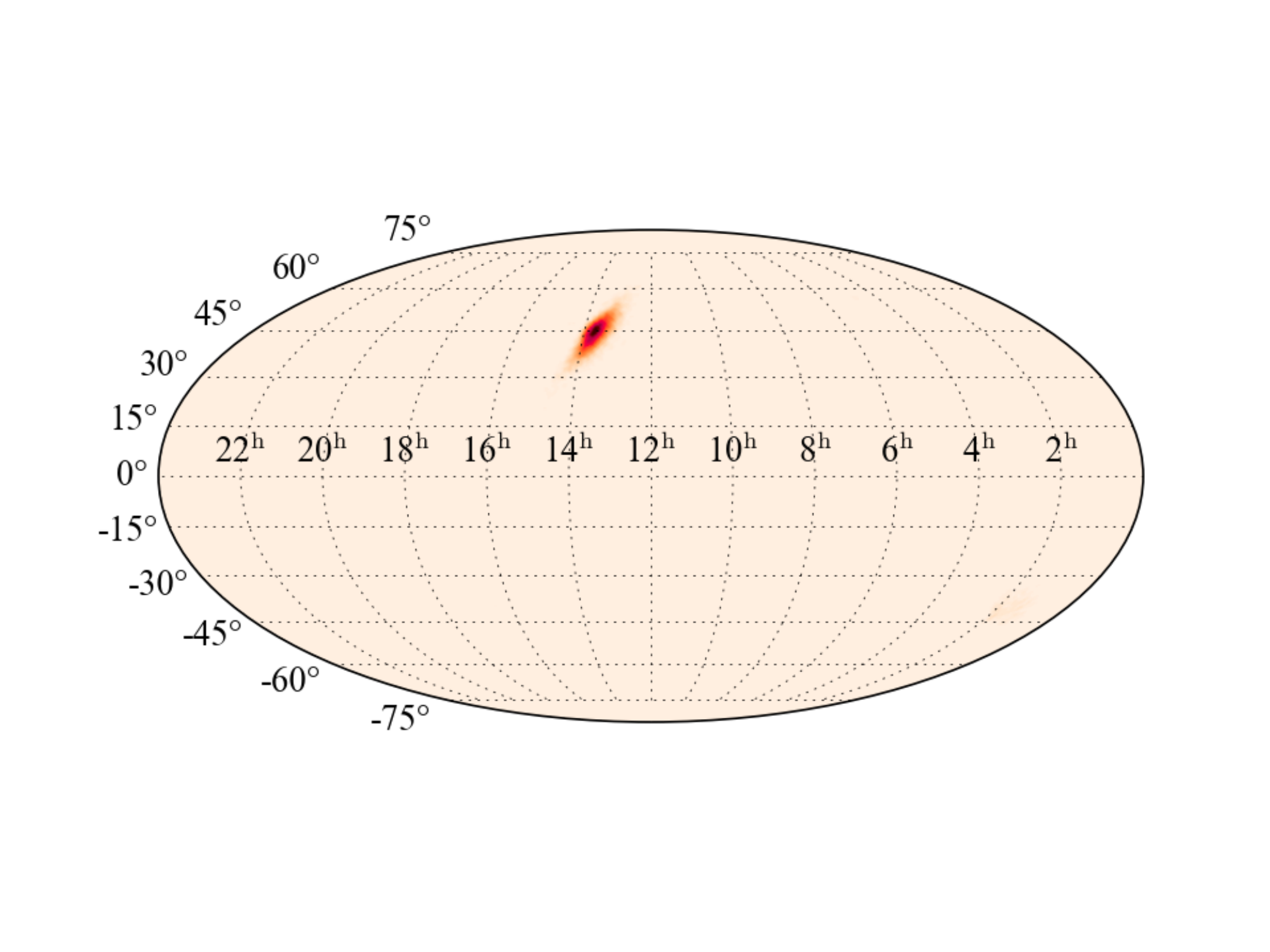}
\end{center}
\end{minipage}
 \begin{minipage}[b]{0.45\linewidth}
 \begin{center}
   \includegraphics[width=1.0\textwidth,angle=0]{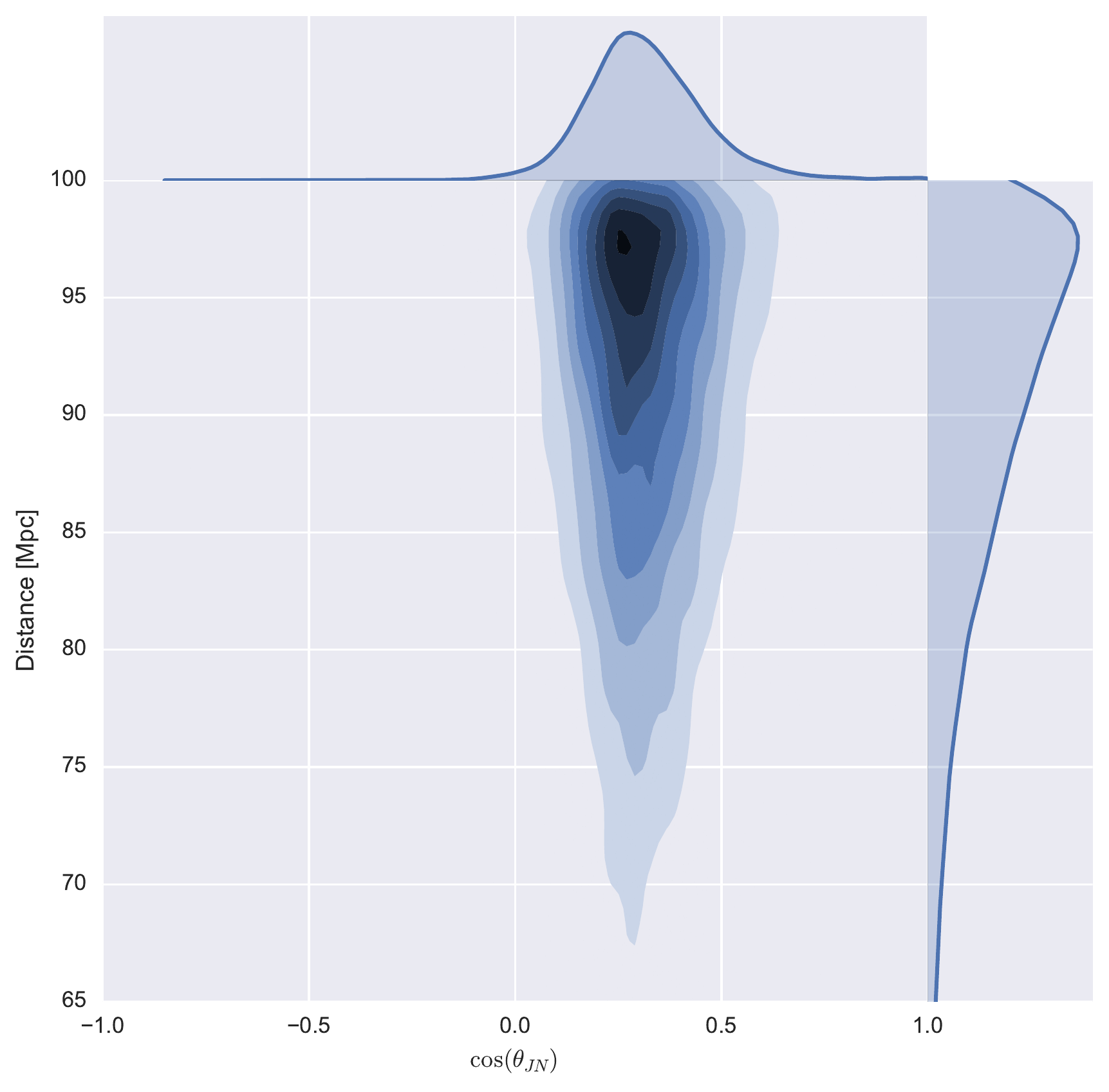}
 \end{center}
 \end{minipage}\\
 \end{tabular}
 \caption{
The same figure as Fig.~\ref{fig:GW150914} but for LVT151012.
}
\label{fig:LVT151012}
 \end{center}
\end{figure}
\begin{figure}[ht]
\begin{center}
\begin{tabular}{cc}
 \begin{minipage}[b]{0.4\linewidth}
 \begin{center}
   \includegraphics[width=1.0\textwidth,angle=0]{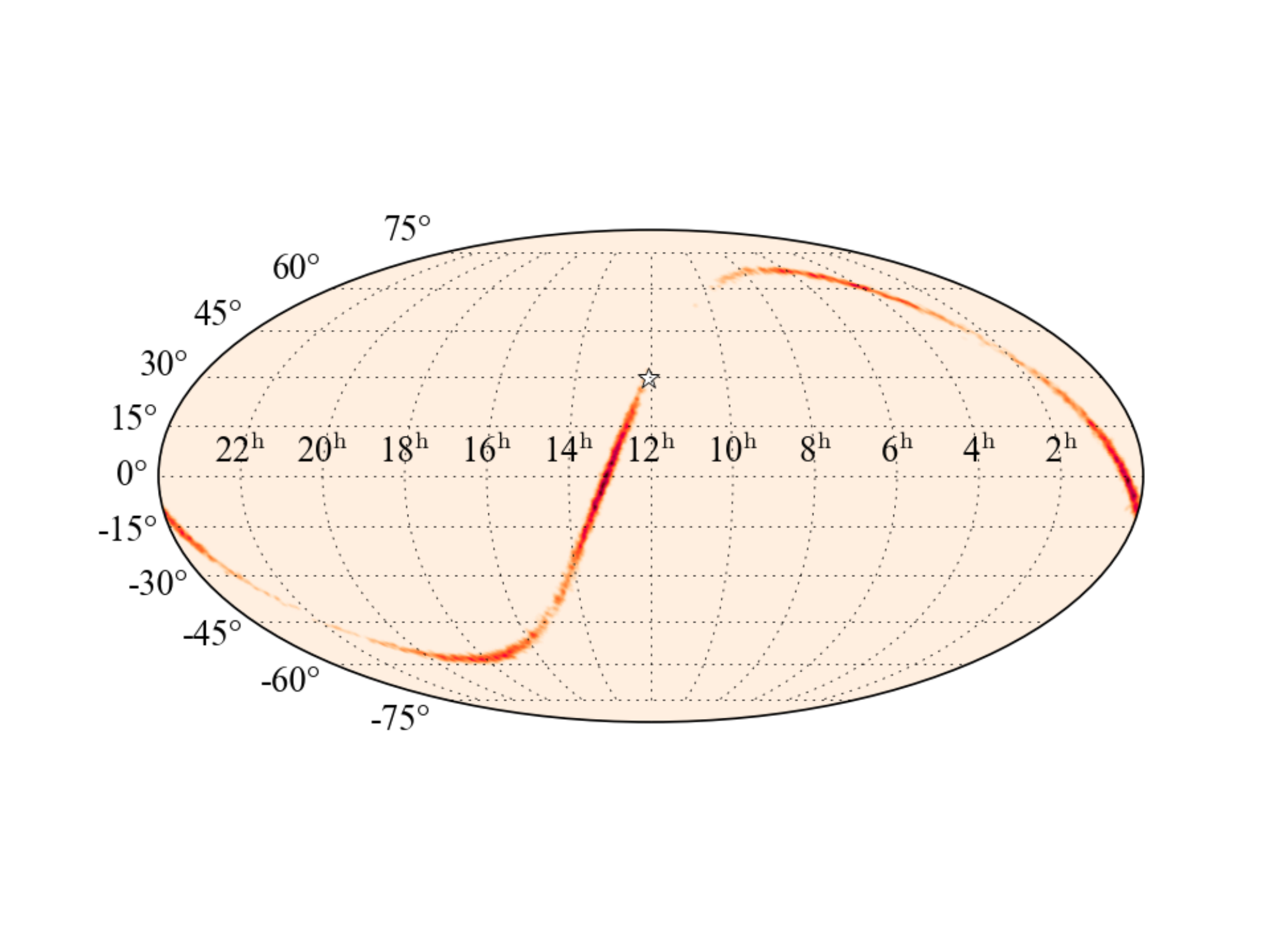}
\end{center}
\end{minipage}
 \begin{minipage}[b]{0.4\linewidth}
 \begin{center}
   \includegraphics[width=1.0\textwidth,angle=0]{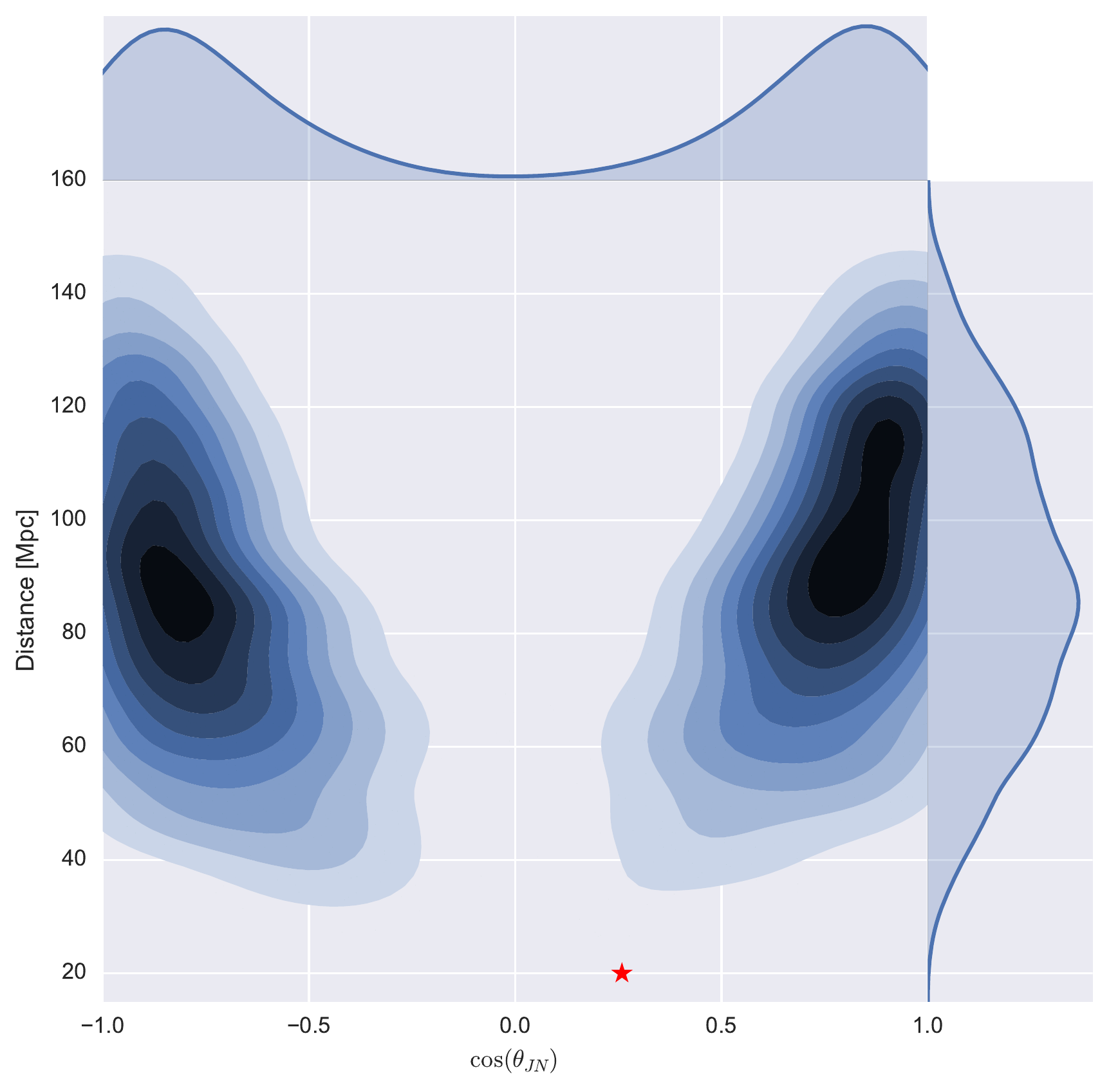}
 \end{center}
 \end{minipage}\\
 \begin{minipage}[b]{0.4\linewidth}
 \begin{center}
   \includegraphics[width=1.0\textwidth,angle=0]{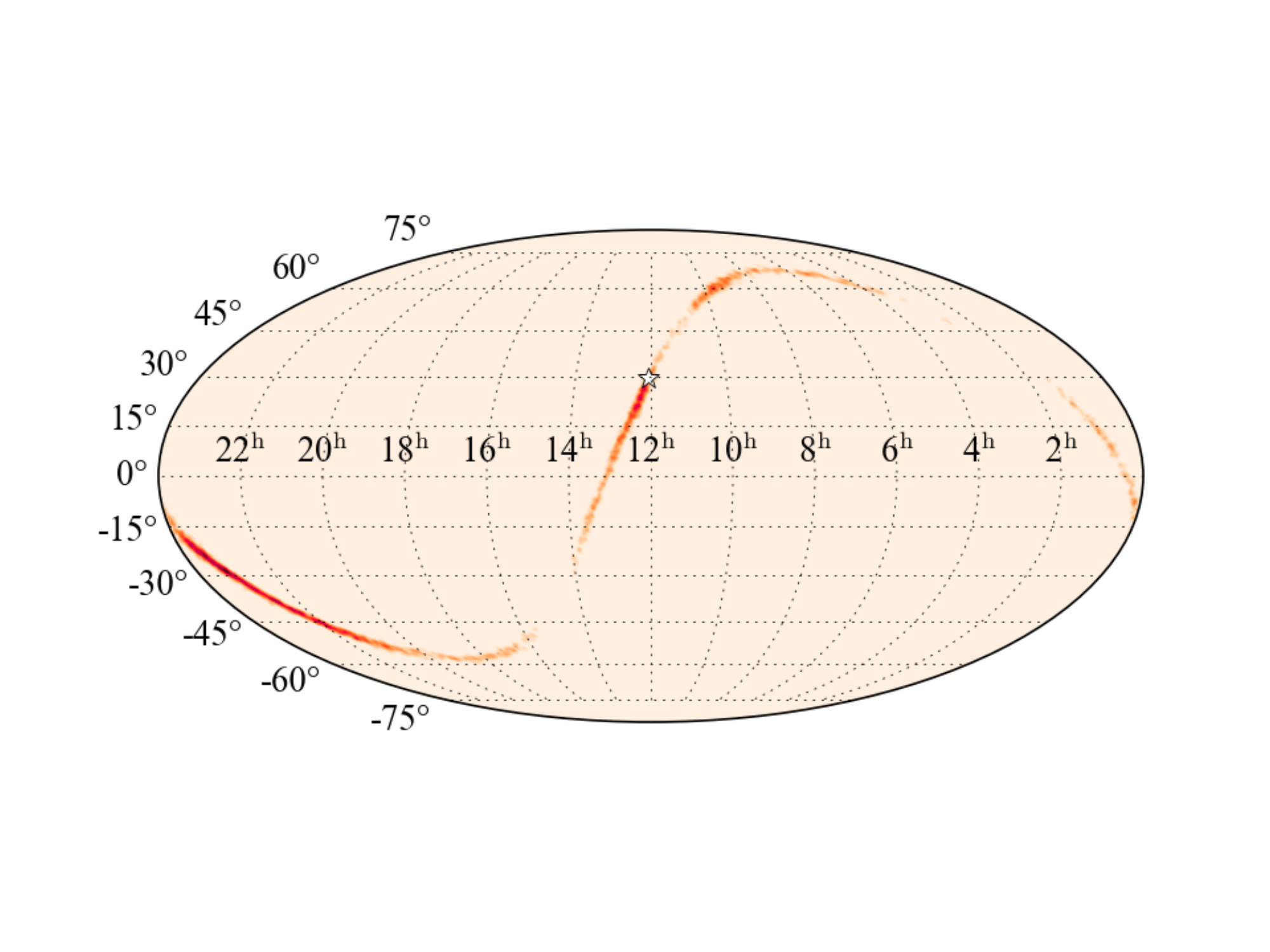}
\end{center}
\end{minipage}
 \begin{minipage}[b]{0.4\linewidth}
 \begin{center}
   \includegraphics[width=1.0\textwidth,angle=0]{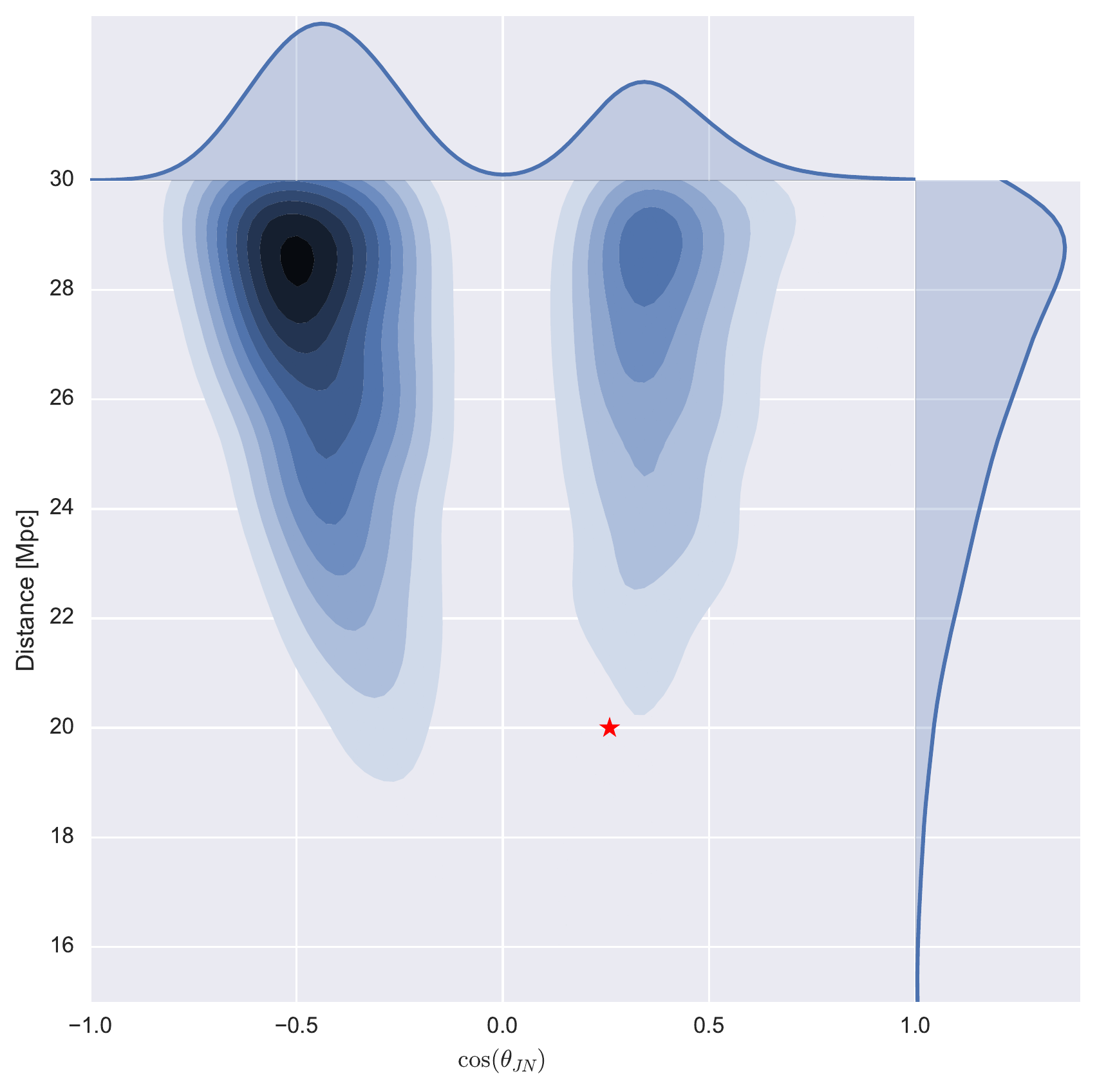}
 \end{center}
 \end{minipage}\\
 \begin{minipage}[b]{0.4\linewidth}
 \begin{center}
   \includegraphics[width=1.0\textwidth,angle=0]{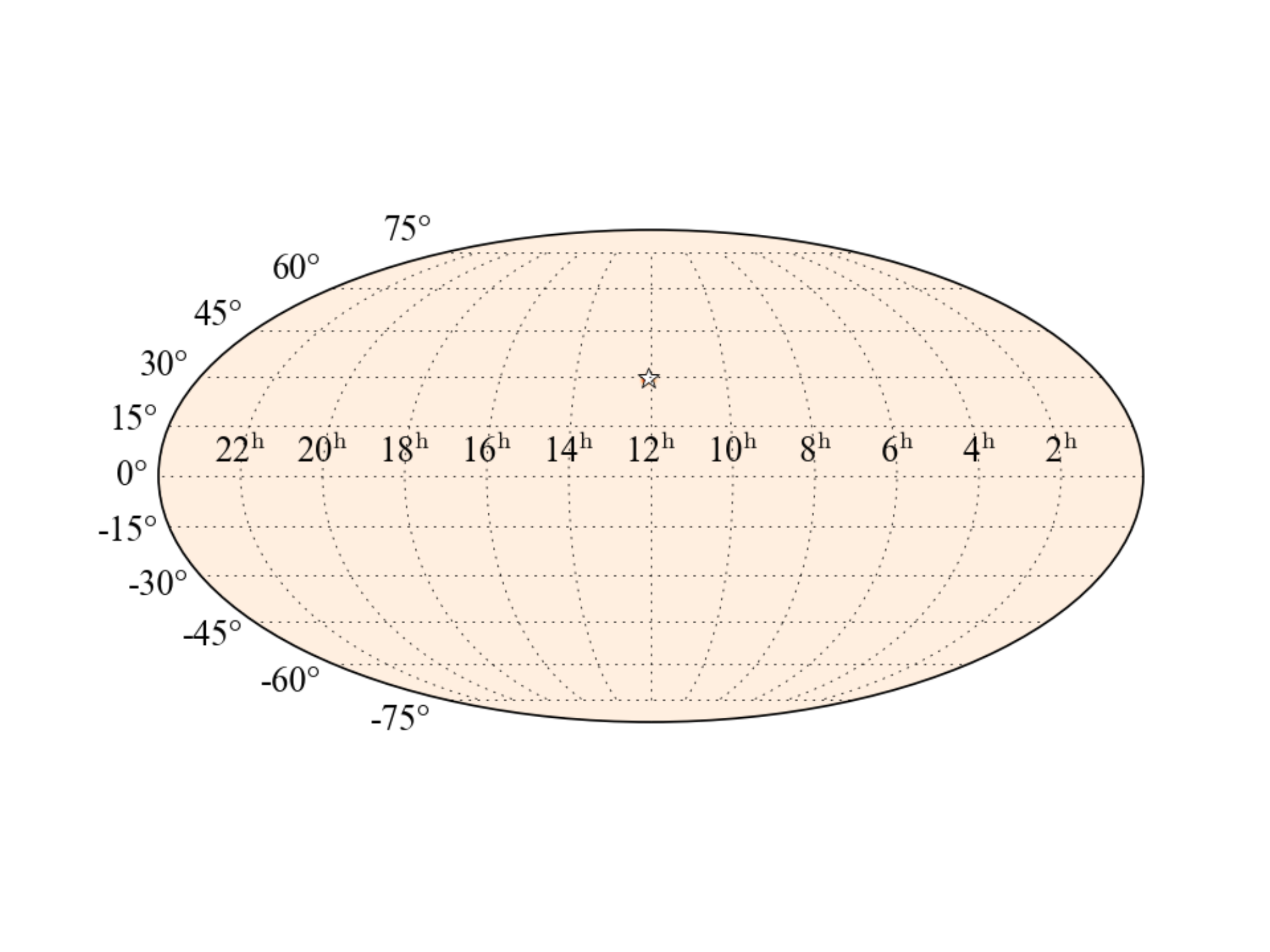}
\end{center}
\end{minipage}
 \begin{minipage}[b]{0.4\linewidth}
 \begin{center}
   \includegraphics[width=1.0\textwidth,angle=0]{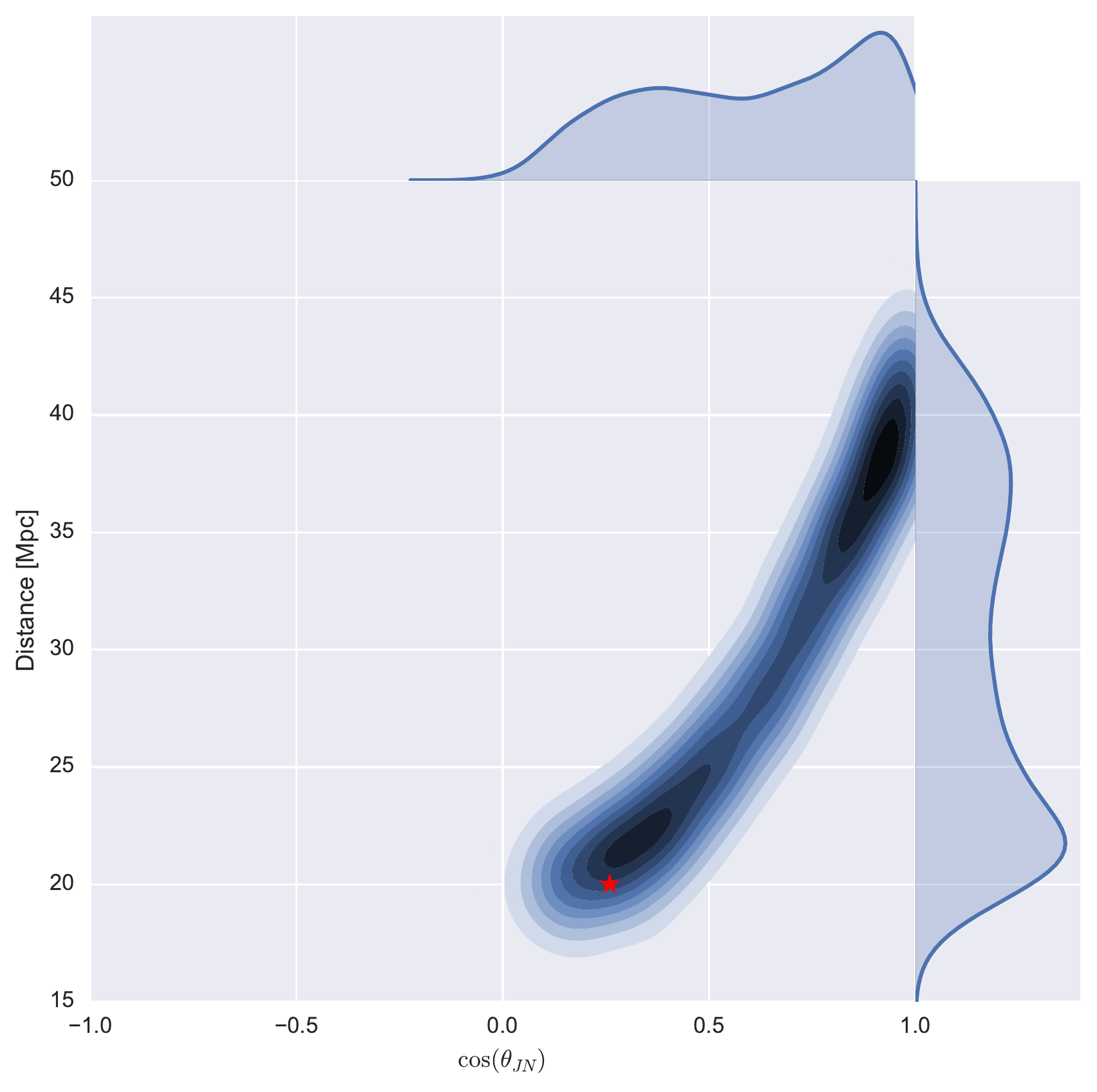}
 \end{center}
 \end{minipage}\\
 \end{tabular}
 \caption{
The posterior PDF for the sky location, and the two dimensional posterior PDF for 
the source luminosity distance and the binary inclination angle 
for a nearly edge-on signal 
with $\theta_{\rm JN}=75$ deg, and $d_{\rm L}=20$Mpc.  
Gaussian noise with LIGO O2 sensitivity is used. 
Results of three cases are shown. 
Case A (top panels): two detectors' network of LIGO Hanford and LIGO Livingston, and no distance prior is used. 
Case B (middle panels): two detectors' network of LIGO Hanford and LIGO Livingston, and a distance prior is set to $d_{\rm L} \leq 30$~Mpc.
Case C (bottom panels): three detectors' network of LIGO Hanford, LIGO Livingston and Virgo, and 
no distance prior is used. 
The star in each figure denotes the location of the injected signal.
}
\label{fig:O2_GoodForVirgo}
 \end{center}
\end{figure}
\begin{figure}[ht]
\begin{center}
\begin{tabular}{cc}
 \begin{minipage}[b]{0.45\linewidth}
 \begin{center}
   \includegraphics[width=1.0\textwidth,angle=0]{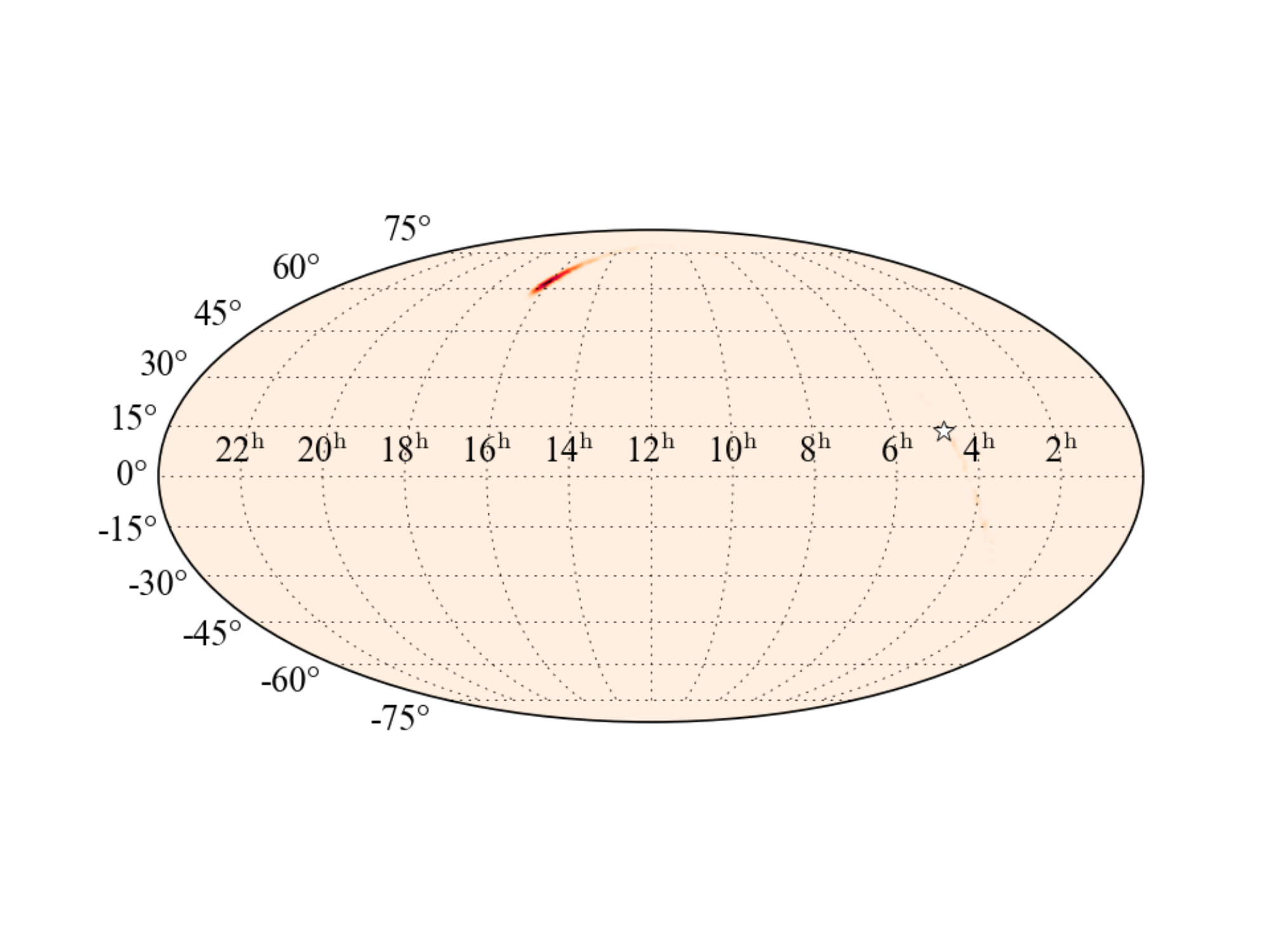}
\end{center}
\end{minipage}
 \begin{minipage}[b]{0.45\linewidth}
 \begin{center}
   \includegraphics[width=1.0\textwidth,angle=0]{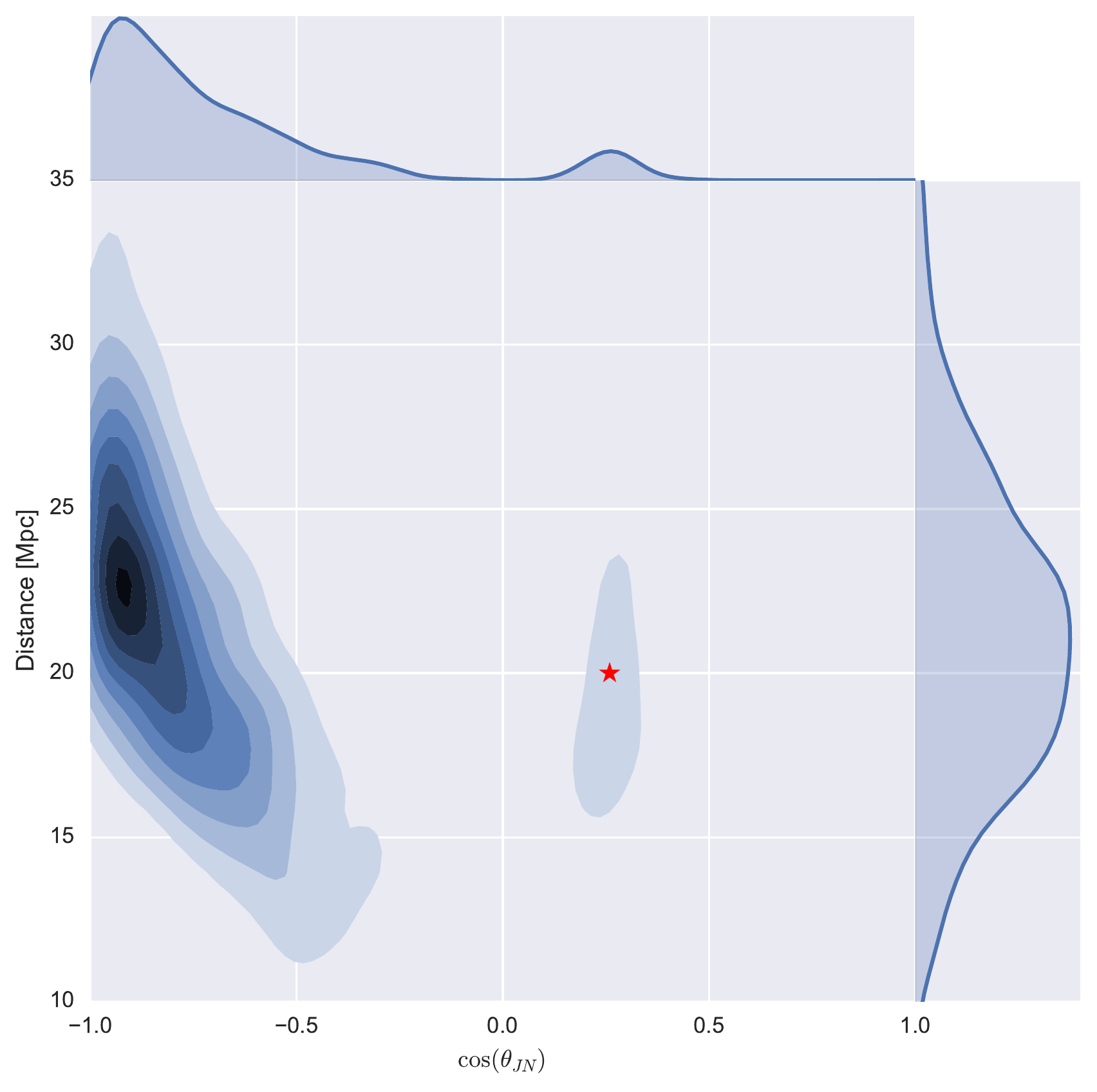}
 \end{center}
 \end{minipage}\\
 \begin{minipage}[b]{0.45\linewidth}
 \begin{center}
   \includegraphics[width=1.0\textwidth,angle=0]{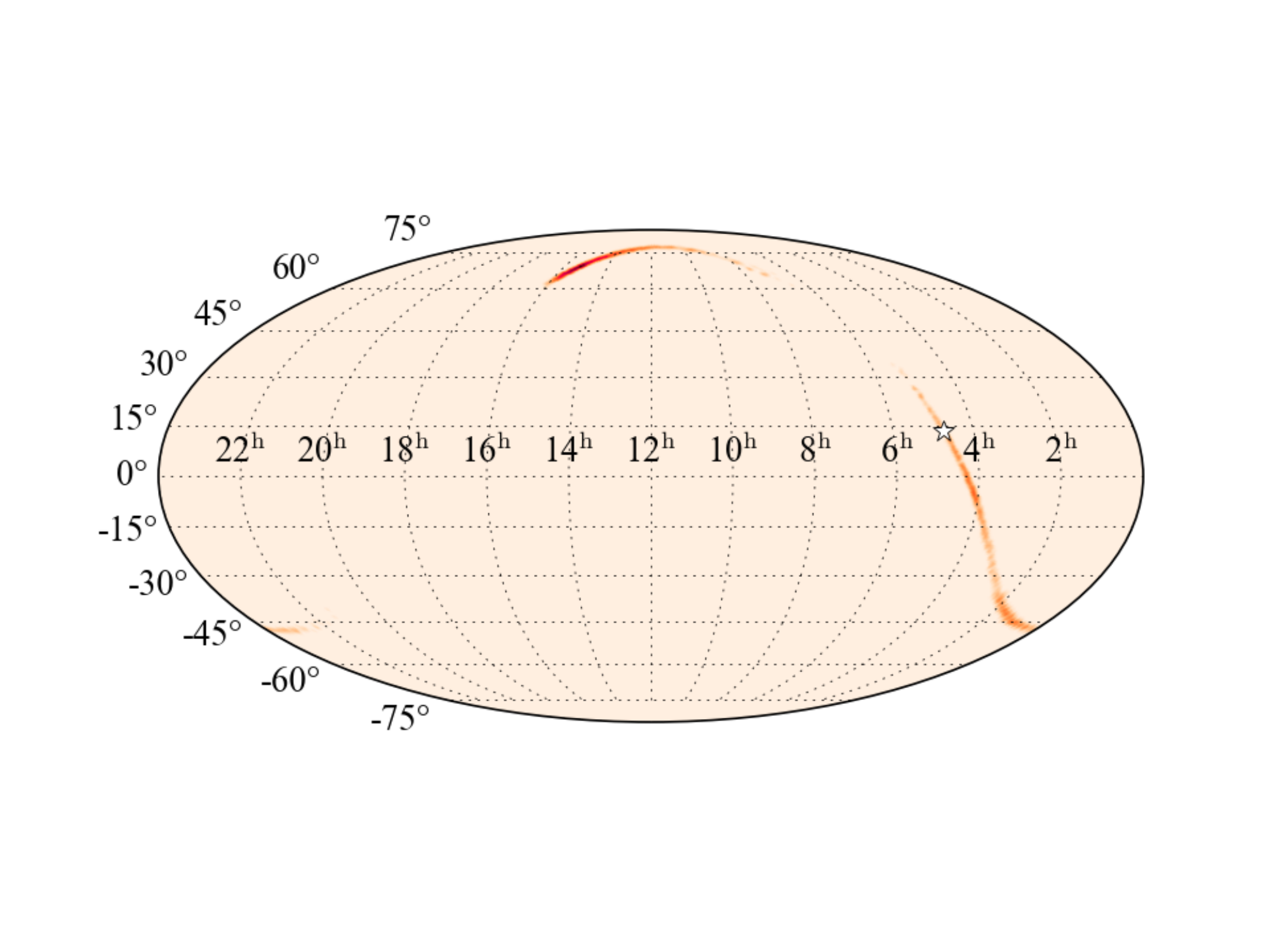}
\end{center}
\end{minipage}
 \begin{minipage}[b]{0.45\linewidth}
 \begin{center}
   \includegraphics[width=1.0\textwidth,angle=0]{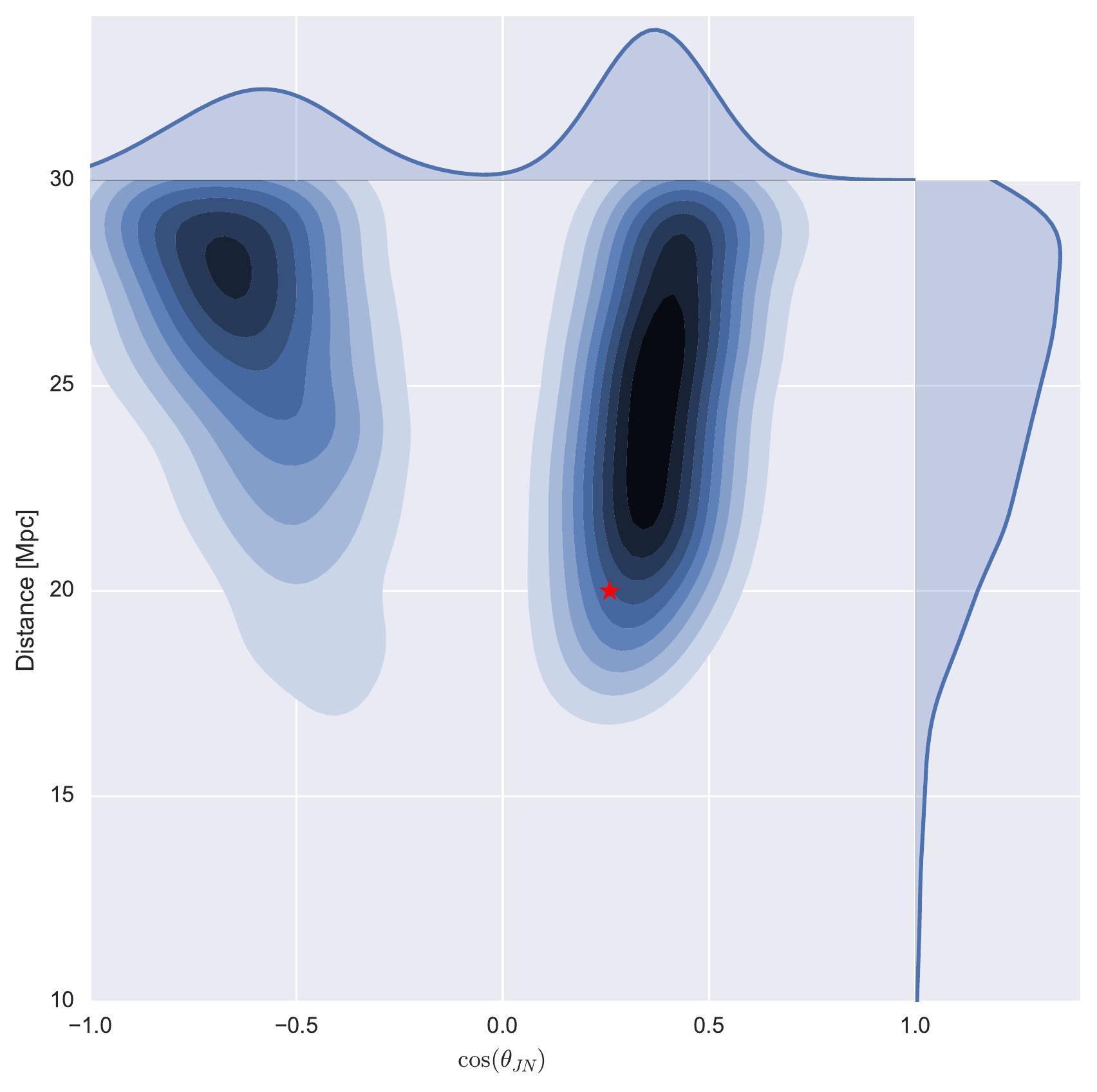}
 \end{center}
 \end{minipage}\\
 \begin{minipage}[b]{0.45\linewidth}
 \begin{center}
   \includegraphics[width=1.0\textwidth,angle=0]{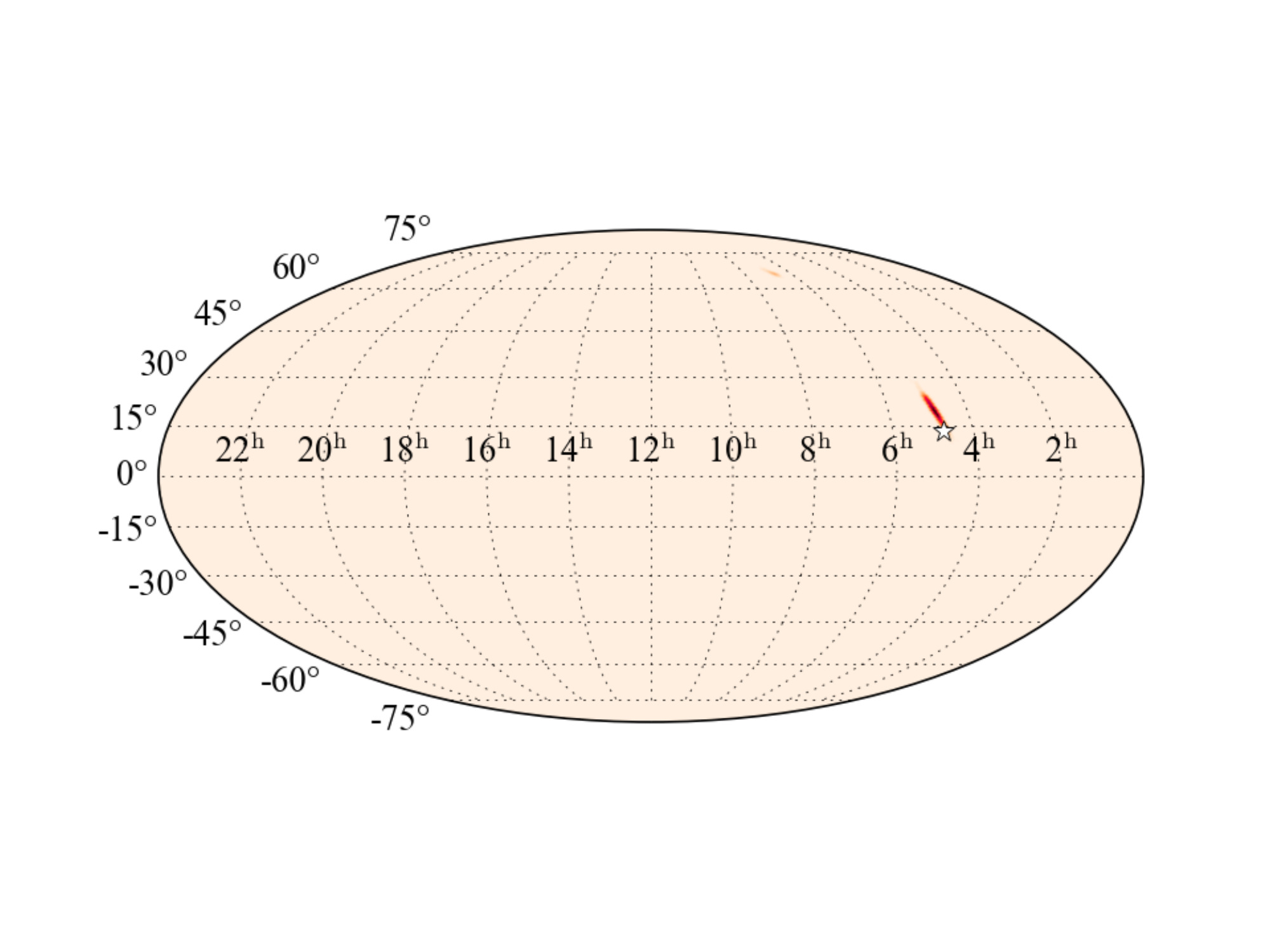}
\end{center}
\end{minipage}
 \begin{minipage}[b]{0.45\linewidth}
 \begin{center}
   \includegraphics[width=1.0\textwidth,angle=0]{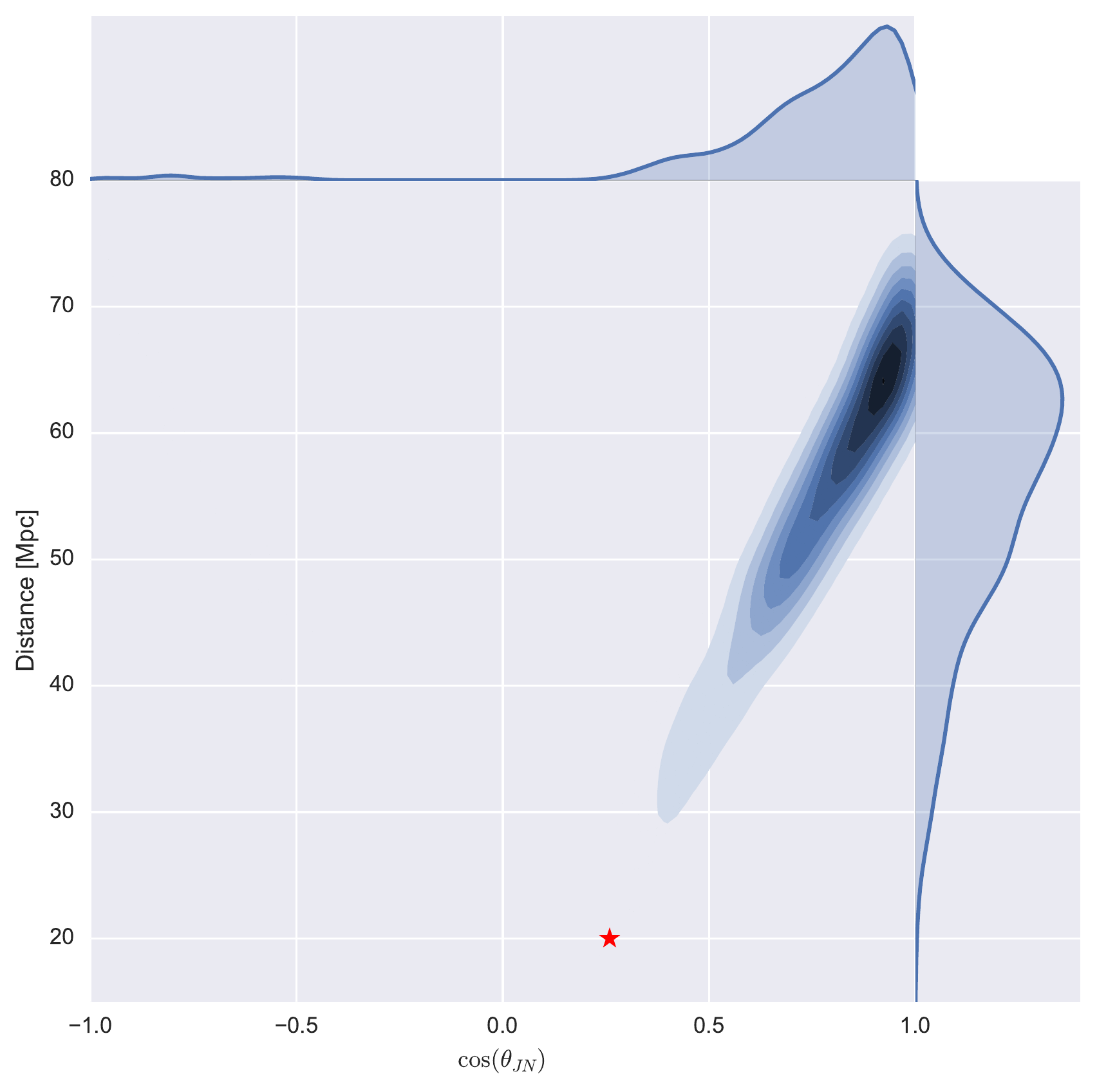}
 \end{center}
 \end{minipage}\\
 \end{tabular}
 \caption{
The same figure as Fig.~\ref{fig:O2_GoodForVirgo} but for a injection signal from different direction.
}
\label{fig:O2_GoodForHL}
 \end{center}
\end{figure}
\begin{figure}[ht]
\begin{center}
\begin{tabular}{cc}
 \begin{minipage}[b]{0.45\linewidth}
 \begin{center}
   \includegraphics[width=1.0\textwidth,angle=0]{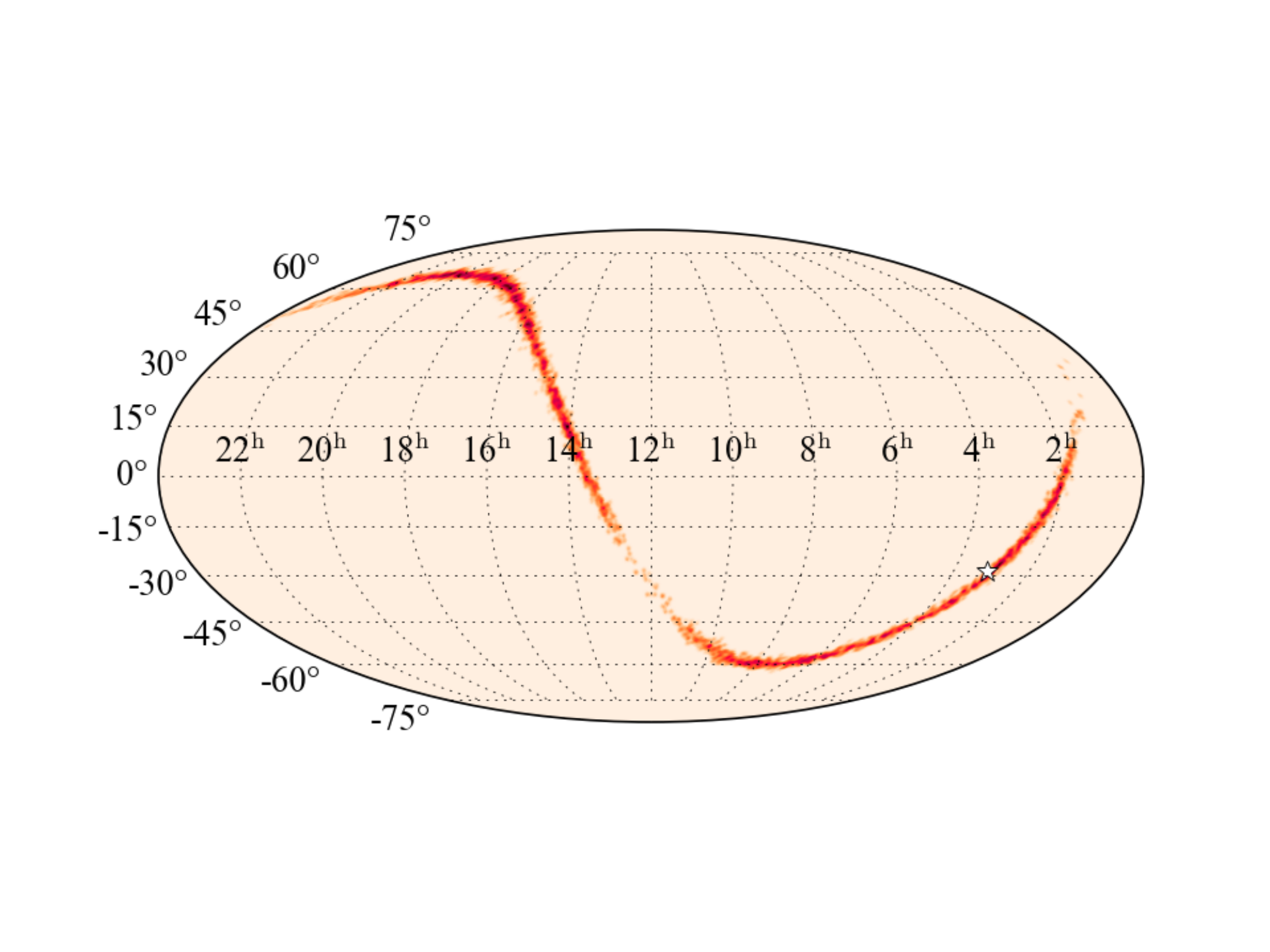}
\end{center}
\end{minipage}
 \begin{minipage}[b]{0.45\linewidth}
 \begin{center}
   \includegraphics[width=1.0\textwidth,angle=0]{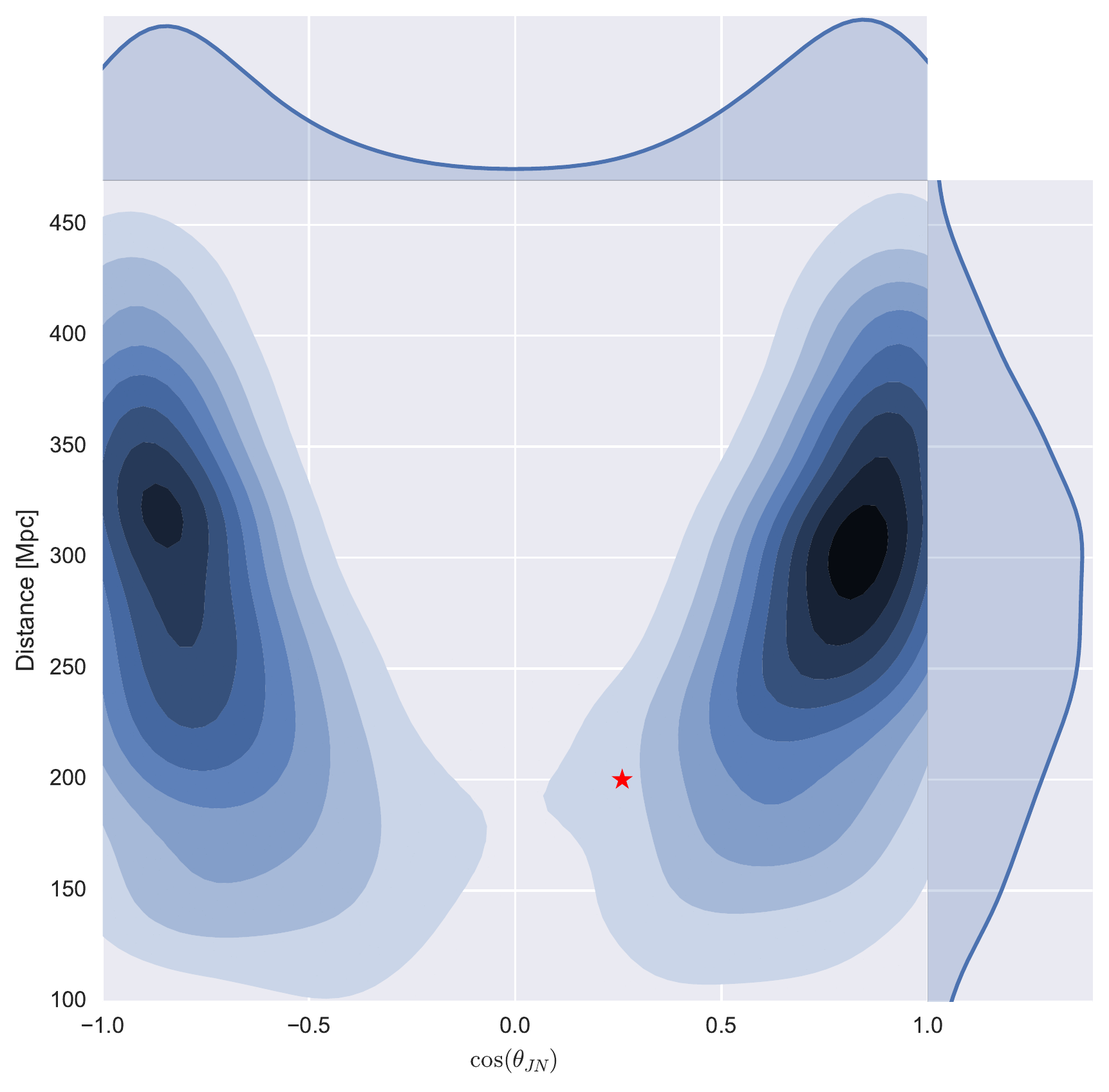}
 \end{center}
 \end{minipage}\\
 \begin{minipage}[b]{0.45\linewidth}
 \begin{center}
   \includegraphics[width=1.0\textwidth,angle=0]{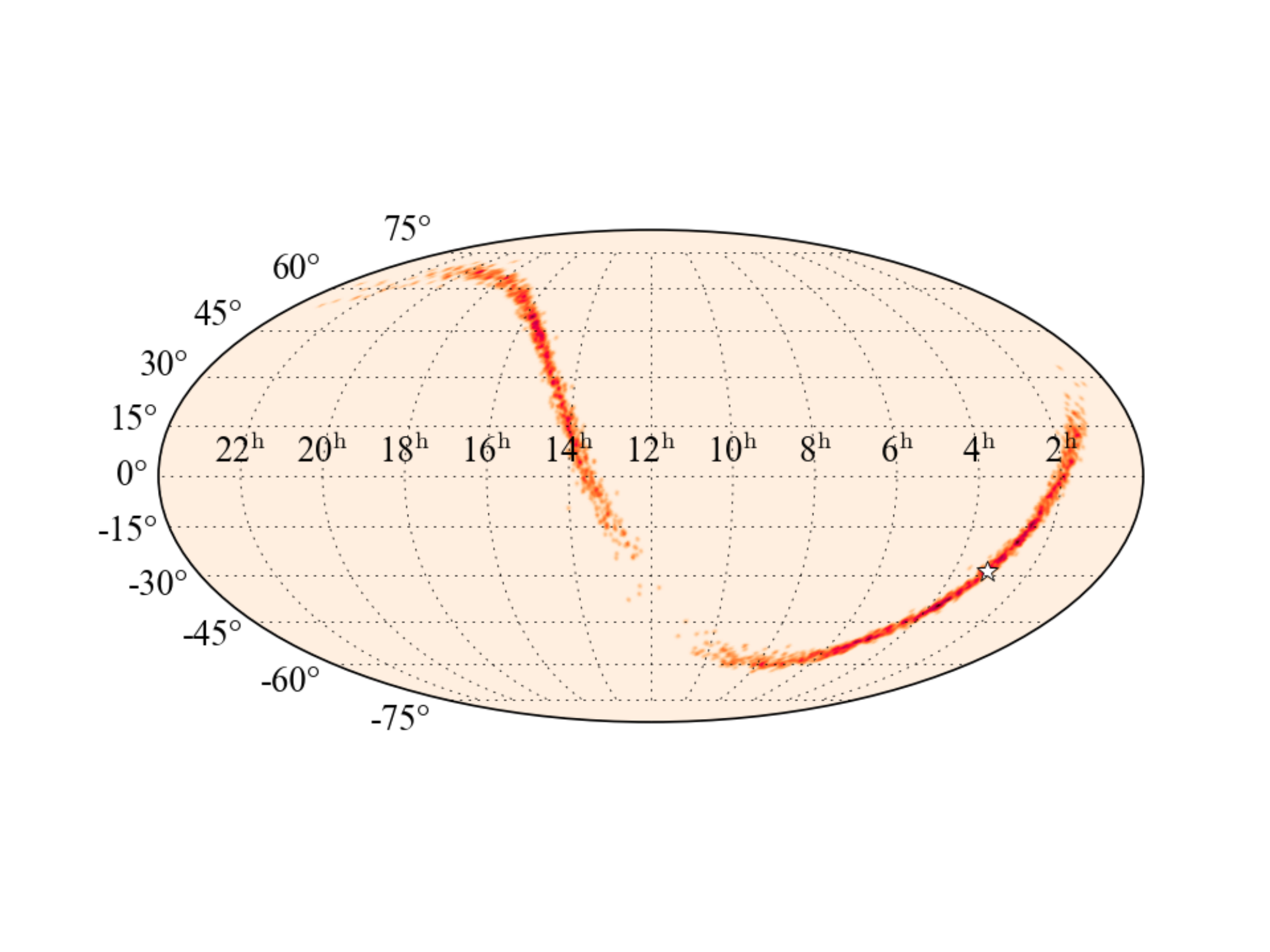}
\end{center}
\end{minipage}
 \begin{minipage}[b]{0.45\linewidth}
 \begin{center}
   \includegraphics[width=1.0\textwidth,angle=0]{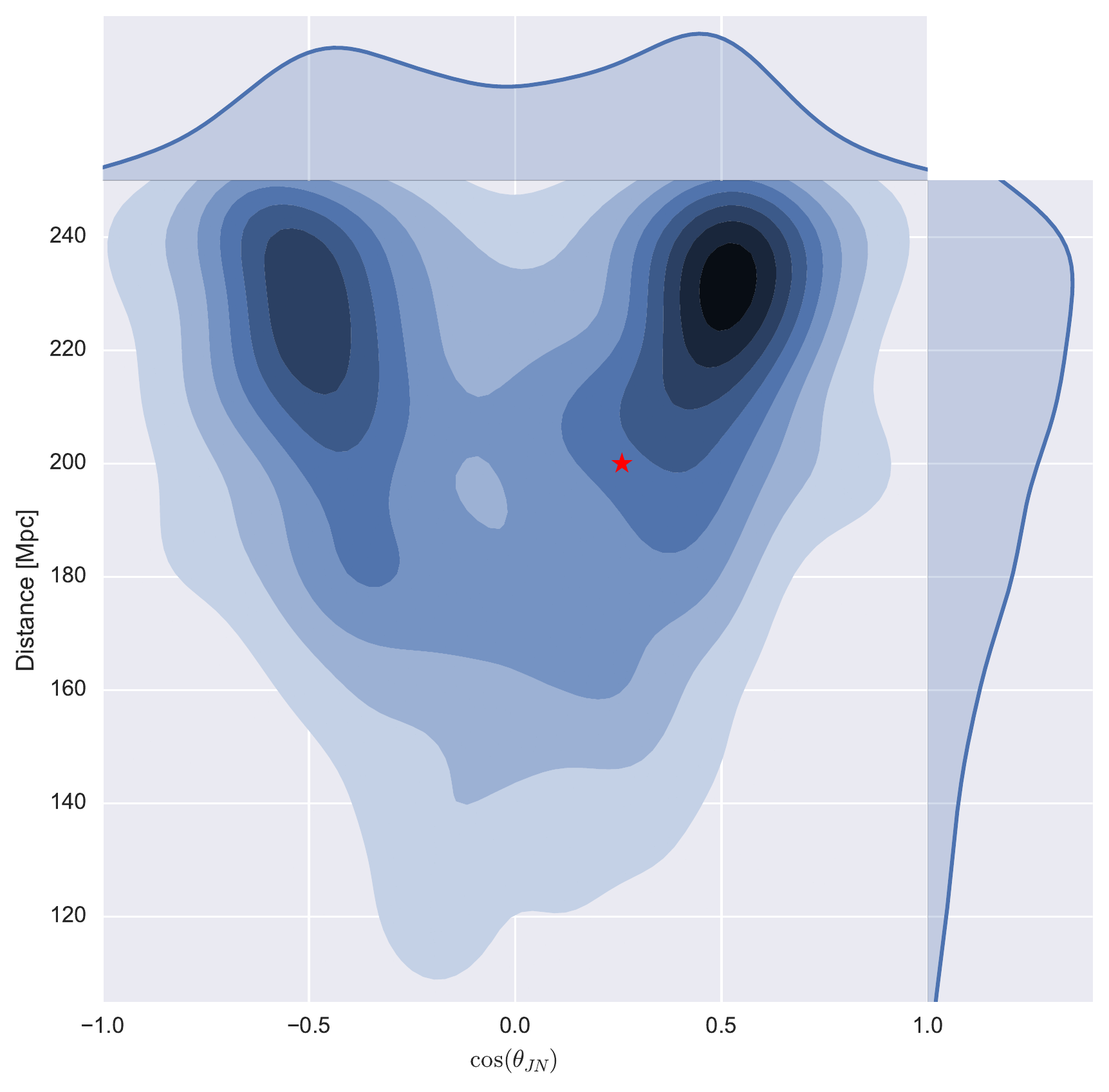}
 \end{center}
 \end{minipage}\\
 \begin{minipage}[b]{0.45\linewidth}
 \begin{center}
   \includegraphics[width=1.0\textwidth,angle=0]{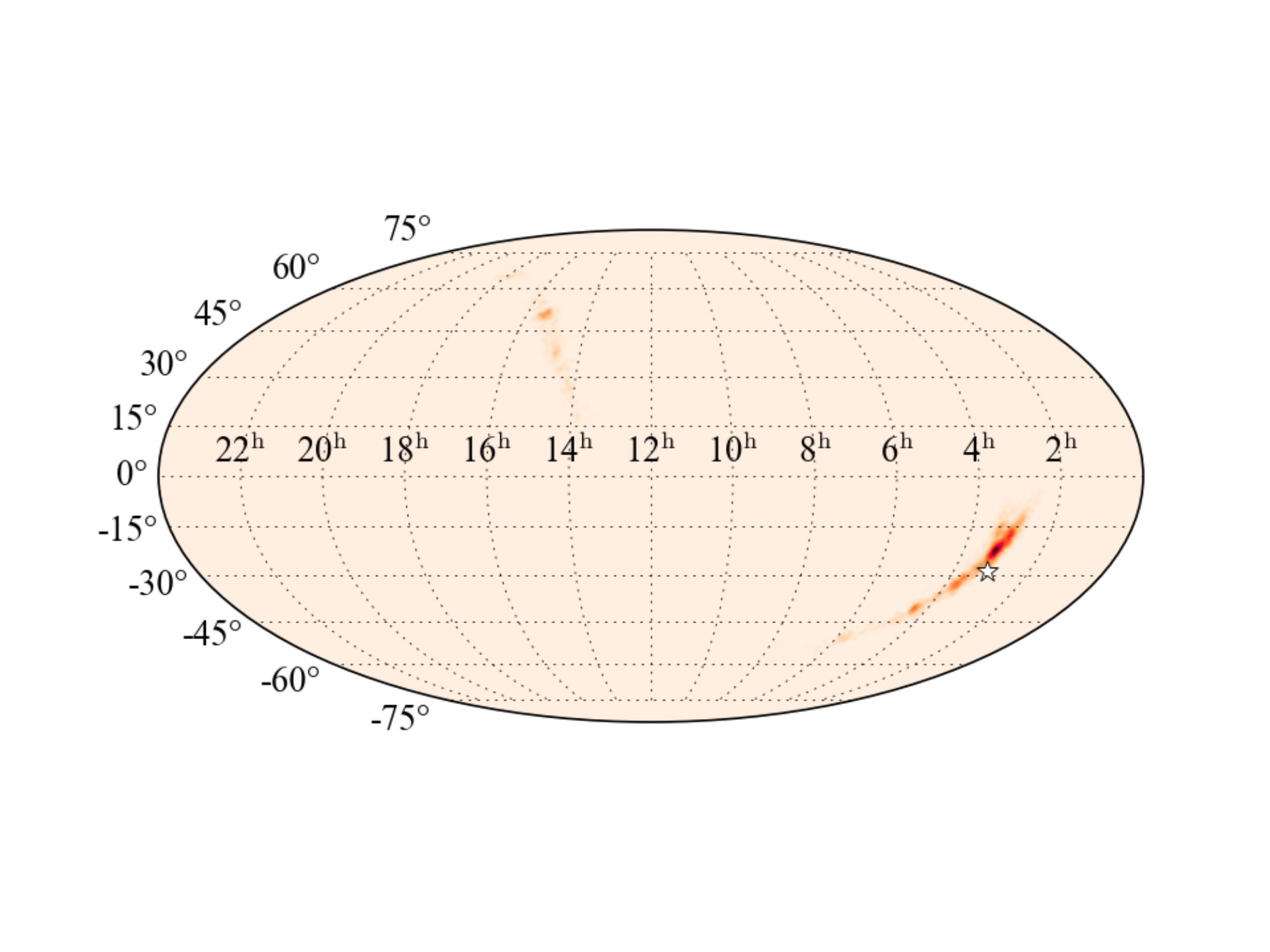}
\end{center}
\end{minipage}
 \begin{minipage}[b]{0.45\linewidth}
 \begin{center}
   \includegraphics[width=1.0\textwidth,angle=0]{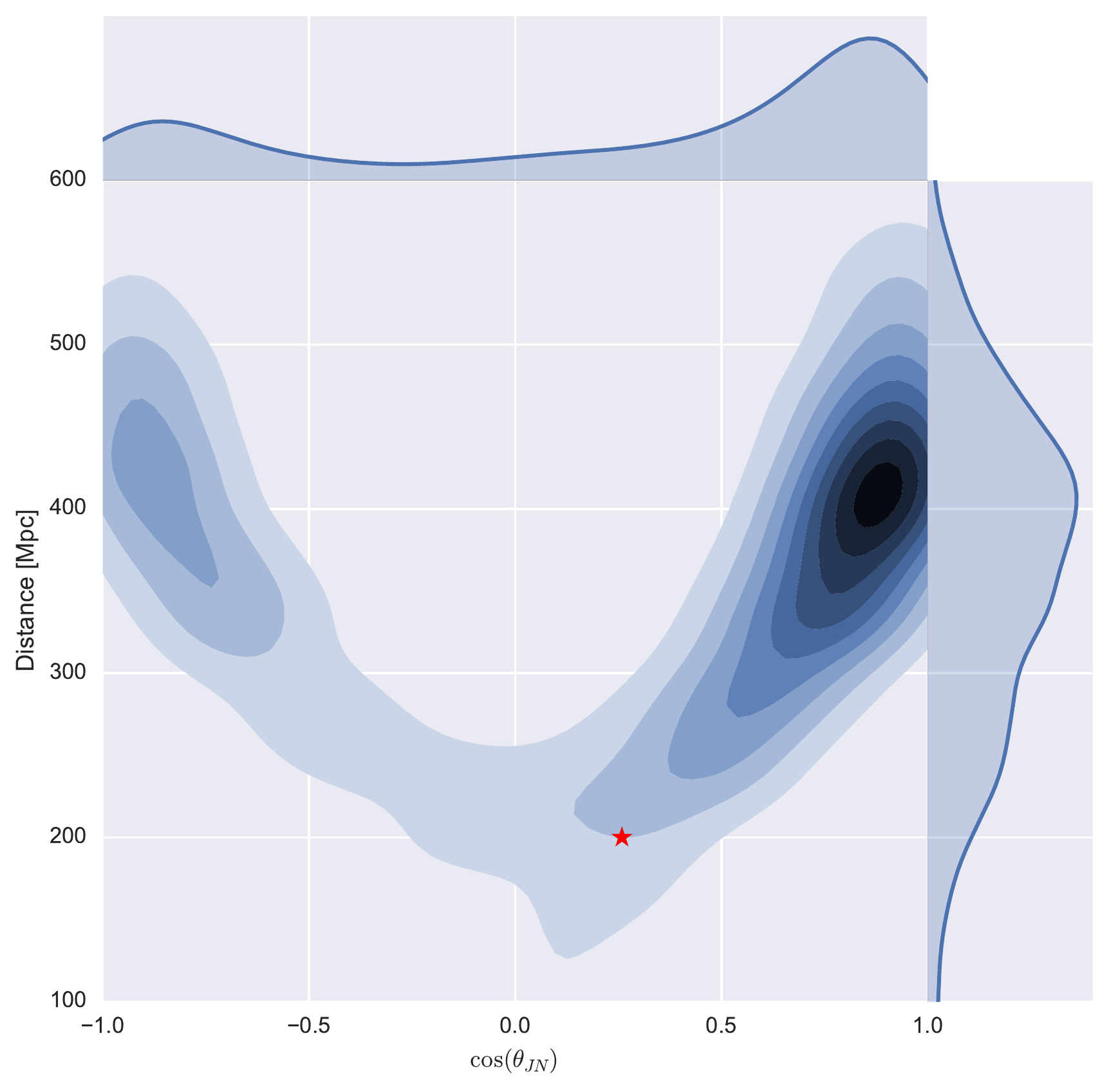}
 \end{center}
 \end{minipage}\\
 \end{tabular}
 \caption{
The same figure as Fig.~\ref{fig:O2_GoodForVirgo} but for a injection signal from different direction,
and for a nearly edge-on signal with $\theta_{\rm JN}=75~{\rm deg}$ and $d_{\rm L}=200~{\rm Mpc}$.
Gaussian noise with the design sensitivities of LIGO and Virgo are used. 
Case B (middle panels): two detectors’ network of LIGO Hanford and LIGO Livingston, and a distance prior is set to $d_{\rm L}\leq 250$~Mpc.
}
\label{fig:event10_BNS200Mpci75deg_DesignHLV}
 \end{center}
\end{figure}
\end{widetext}


\begin{thebibliography}{99}
\bibitem{Abbott:2016blz} 
  B.~P.~Abbott {\it et al.}, 
  Observation of Gravitational Waves from a Binary Black Hole Merger,
  Phys.\ Rev.\ Lett.\  {\bf 116}, 061102 (2016)
  [arXiv:1602.03837 [gr-qc]].

\bibitem{TheLIGOScientific:2016pea} 
  B.~P.~Abbott {\it et al.}, 
  Binary Black Hole Mergers in the first Advanced LIGO Observing Run,
  Phys.\ Rev.\ X {\bf 6}, 041015 (2016)
  [arXiv:1606.04856 [gr-qc]].
    
\bibitem{Abbott:2016gcq} 
  B.~P.~Abbott {\it et al.}, 
  Localization and broadband follow-up of the gravitational-wave transient GW150914,
  Astrophys.\ J.\  {\bf 826}, L13 (2016)
  [arXiv:1602.08492 [astro-ph.HE]].

\bibitem{Abbott:2016iqz} 
  B.~P.~Abbott {\it et al.}, 
  Supplement: Localization and broadband follow-up of the gravitational-wave transient GW150914,
  Astrophys.\ J.\ Suppl.\  {\bf 225}, 8 (2016)
  [arXiv:1604.07864 [astro-ph.HE]].
              
\bibitem{Morokuma:2016hqx} 
  T.~Morokuma {\it et al.},
  J-GEM Follow-Up Observations to Search for an Optical Counterpart of The First Gravitational Wave Source GW150914,
  Publ.\ Astron.\ Soc.\ Jap.\  {\bf 68}, L9 (2016)
  [arXiv:1605.03216 [astro-ph.SR]].
 
\bibitem{Smartt:2016oeu} 
  S.~J.~Smartt {\it et al.},
  A search for an optical counterpart to the gravitational wave event GW151226,
  Astrophys.\ J.\  {\bf 827}, L40 (2016)
  [arXiv:1606.04795 [astro-ph.HE]].

\bibitem{Yoshida:2016ddu} 
  M.~Yoshida {\it et al.},
  J-GEM Follow-Up Observations of The Gravitational Wave Source GW151226,
  Publ.\ Astron.\ Soc.\ Jap.\  {\bf 69}, 9 (2017)
  [arXiv:1611.01588 [astro-ph.HE]].

\bibitem{ref:Fermi}
Note however, 
  V. Connaughton {\it et al.}, 
  Fermi GBM observations of ligo gravitational-wave event GW150914,
  Astrophys.\ J.\ Lett.,\  {\bf 826}, L6 (2016).

\bibitem{Tanaka:2016sbx} 
  M.~Tanaka,
  Kilonova/Macronova Emission from Compact Binary Mergers,
  Adv.\ Astron.\  {\bf 2016}, 6341974 (2016)
  [arXiv:1605.07235 [astro-ph.HE]].
  
\bibitem{Schutz:2011tw} 
  B.~F.~Schutz,
  Networks of gravitational wave detectors and three figures of merit,
  Class.\ Quant.\ Grav.\  {\bf 28}, 125023 (2011)
  [arXiv:1102.5421 [astro-ph.IM]].

\bibitem{Nissanke:2012dj} 
  S.~Nissanke, M.~Kasliwal and A.~Georgieva,
  Identifying Elusive Electromagnetic Counterparts to Gravitational Wave Mergers: an end-to-end simulation,
  Astrophys.\ J.\  {\bf 767}, 124 (2013)
  [arXiv:1210.6362 [astro-ph.HE]].

\bibitem{Rover:2007ij} 
  C.~Rover, R.~Meyer, G.~M.~Guidi, A.~Vicere and N.~Christensen,
  Coherent Bayesian analysis of inspiral signals,
  Class.\ Quant.\ Grav.\  {\bf 24}, S607 (2007)
  [arXiv:0707.3962 [gr-qc]].

\bibitem{Messenger:2012jy} 
  C.~Messenger and J.~Veitch,
  Avoiding selection bias in gravitational wave astronomy,
  New J.\ Phys.\  {\bf 15}, 053027 (2013)
  [arXiv:1206.3461 [astro-ph.IM]].

\bibitem{Salafia:2017ebv} 
  O.~S.~Salafia, M.~Colpi, M.~Branchesi, E.~Chassande-Mottin, G.~Ghirlanda, G.~Ghisellini and S.~Vergani,
  Where and when: optimal scheduling of the electromagnetic follow-up of gravitational-wave events based on counterpart lightcurve models,
  Astrophys.\ J.\ {\bf 846}, 62 (2017)
  [arXiv:1704.05851 [astro-ph.HE]].

\bibitem{Virgo}
 The Virgo Collaboration, Advanced Virgo Baseline Design (2009). (Available at: https://tds.ego-gw.it/ql/?c=6589, date last accessed August 20, 2016).

\bibitem{Somiya:2011np} 
  K.~Somiya et al.,
  Detector configuration of KAGRA: The Japanese cryogenic gravitational-wave detector,
  Class. Quant. Grav. {\bf 29}, 124007 (2012)
  [arXiv:1111.7185 [gr-qc]].
  
\bibitem{Aso:2013eba} 
  Y.~Aso, Y.~Michimura, K.~Somiya, M.~Ando, O.~Miyakawa, T.~Sekiguchi, D.~Tatsumi, and H.~Yamamoto,
  et al., 
  Interferometer design of the KAGRA gravitational wave detector,
  Phys.\ Rev.\ D {\bf 88}, 043007 (2013)
  [arXiv:1306.6747 [gr-qc]].

\bibitem{events}
  LIGO Scientific Collaboration, LIGO Open Science Center release of GW150914, LVT151012, GW151226, 2016, 
DOI 10.7935/K5MW2F23, 10.7935/K5CC0XMZ, 10.7935/K5H41PBP.


\bibitem{Skilling:2006}
 J.~Skilling, 
 Bayesian Analysis {\bf 1}, 833 (2006).

\bibitem{Veitch:2009hd} 
  J.~Veitch and A.~Vecchio,
  Bayesian coherent analysis of in-spiral gravitational wave signals with a detector network,
  Phys.\ Rev.\ D {\bf 81}, 062003 (2010)
  [arXiv:0911.3820 [astro-ph.CO]].

\bibitem{Veitch:2014wba} 
  J.~Veitch {\it et al.},
  Parameter estimation for compact binaries with ground-based gravitational-wave observations using the LALInference software library,
  Phys.\ Rev.\ D {\bf 91}, 042003 (2015)
  [arXiv:1409.7215 [gr-qc]].

\bibitem{TheLIGOScientific:2016wfe} 
  B.~P.~Abbott {\it et al.}, 
  Properties of the Binary Black Hole Merger GW150914,
  Phys.\ Rev.\ Lett.\  {\bf 116}, 241102 (2016)
  [arXiv:1602.03840 [gr-qc]].  

\bibitem{Hannam:2013oca} 
  M.~Hannam, P.~Schmidt, A.~Boh$\acute{e}$, L.~Haegel, S.~Husa, F.~Ohme, G.~Pratten and M.~P$\ddot{u}$rrer,
  Simple Model of Complete Precessing Black-Hole-Binary Gravitational Waveforms,
  Phys.\ Rev.\ Lett.\  {\bf 113}, 151101 (2014)
  [arXiv:1308.3271 [gr-qc]].
  
\bibitem{Schmidt:2012rh} 
  P.~Schmidt, M.~Hannam and S.~Husa,
  Towards models of gravitational waveforms from generic binaries: A simple approximate mapping between precessing and non-precessing inspiral signals,
  Phys.\ Rev.\ D {\bf 86}, 104063 (2012)
  [arXiv:1207.3088 [gr-qc]].
  
\bibitem{Aasi:2013wya} 
  J.~Aasi {\it et al.}, 
  Prospects for Observing and Localizing Gravitational-Wave Transients with Advanced LIGO and Advanced Virgo,
  Living Revew Relativity  {\bf 19}, 1 (2016)
  [arXiv:1304.0670 [gr-qc]].

\bibitem{Singer:2014qca} 
  L.~P.~Singer {\it et al.},
  The First Two Years of Electromagnetic Follow-Up with Advanced LIGO and Virgo,
  Astrophys.\ J.\  {\bf 795}, 105 (2014)
  [arXiv:1404.5623 [astro-ph.HE]].

\bibitem{Buonanno:2009zt} 
  A.~Buonanno, B.~Iyer, E.~Ochsner, Y.~Pan and B.~S.~Sathyaprakash,
  Comparison of post-Newtonian templates for compact binary inspiral signals in gravitational-wave detectors,
  Phys.\ Rev.\ D {\bf 80}, 084043 (2009)
  [arXiv:0907.0700 [gr-qc]].

\bibitem{Singer:2016eax} 
  L.~P.~Singer {\it et al.},
  Going the Distance: Mapping Host Galaxies of LIGO and Virgo Sources in Three Dimensions Using Local Cosmography and Targeted Follow-up,
  Astrophys.\ J.\  {\bf 829}, L15 (2016)
  [arXiv:1603.07333 [astro-ph.HE]].
  
        
\end{thebibliography}
\end{document}